\documentclass[11pt]{article}

\usepackage[utf8]{inputenc}
\usepackage[left=25mm,right=25mm,top=25mm,bottom=25mm]{geometry}
\usepackage{latexsym}
\usepackage{graphicx}
\usepackage{multicol,multirow}
\usepackage{amsmath,amssymb,amsfonts}
\usepackage{mathrsfs}
\usepackage{amsthm}
\usepackage{rotating}
\usepackage{appendix}
\usepackage{ifpdf}
\usepackage[T1]{fontenc}
\usepackage{times}
\usepackage{sourcesanspro}
\usepackage{newtxmath}
\usepackage{textcomp}%
\usepackage{xcolor}%
\usepackage{authblk}

\usepackage{booktabs}

\usepackage{hyperref}

\usepackage{tikz}
\usepackage{pgfplots}
\usepgfplotslibrary{groupplots}
\usepackage{pgfplotstable}
\usepackage{subcaption}
\usepackage[capitalise]{cleveref}
\usepackage{siunitx}
 
\usepackage{mathtools} % for \mathllap

\theoremstyle{definition}
\newtheorem{definition}{Definition}%[section]
\newtheorem{prop}{Proposition}%[section]
\pgfplotsset{compat=1.16}
\usepgfplotslibrary{fillbetween}
\usetikzlibrary{patterns, fit, backgrounds}
\usepgfplotslibrary{colormaps}

\usepackage[maxbibnames=20, maxcitenames=2, 
    natbib=true, style=numeric,
    isbn=false, url=false,giveninits=true,sortcites]{biblatex}   % Literaturverzeichnis
\bibliography{bibliography}
\definecolor{color11}{RGB}{236, 188, 0}
\definecolor{color12}{RGB}{50, 67, 121}
\definecolor{color13}{RGB}{222, 222, 222}
\definecolor{color21}{RGB}{92, 134, 196}
\definecolor{color22}{RGB}{249, 156, 0}
\definecolor{color23}{RGB}{222, 222, 222}
\definecolor{color31}{RGB}{222, 222, 222}
\definecolor{color32}{RGB}{222, 222, 222}
\definecolor{color33}{RGB}{191, 60, 60}

\pgfplotscreateplotcyclelist{mycolorlist}{%
only marks,solid, mark options={solid,fill=color11,fill opacity=1,mark size=4pt},mark=square*,color11\\
only marks,solid, mark options={solid,fill=color12,fill opacity=1,mark size=4pt},mark=square*,color12\\
only marks,solid, mark options={solid,fill=color13,fill opacity=1,mark size=4pt},mark=square*,color13\\
only marks,solid, mark options={solid,fill=color21,fill opacity=1,mark size=4pt},mark=square*,color21\\
only marks,solid, mark options={solid,fill=color22,fill opacity=1,mark size=4pt},mark=square*,color22\\
only marks,solid, mark options={solid,fill=color23,fill opacity=1,mark size=4pt},mark=square*,color23\\
only marks,solid, mark options={solid,fill=color31,fill opacity=1,mark size=4pt},mark=square*,color31\\
only marks,solid, mark options={solid,fill=color32,fill opacity=1,mark size=4pt},mark=square*,color32\\
only marks,solid, mark options={solid,fill=color33,fill opacity=1,mark size=4pt},mark=square*,color33\\
color11, dashed, very thick,mark=*, mark options={solid}, mark repeat = 15\\
color12, dashed, very thick ,mark=*, mark options={solid}, mark repeat = 15\\
color13, dashed, very thick ,mark=*, mark options={solid}, mark repeat = 15\\
color21, dashed, very thick ,mark=*, mark options={solid}, mark repeat = 15\\
color22, dashed, very thick ,mark=*, mark options={solid}, mark repeat = 15\\
color23, dashed, very thick ,mark=*, mark options={solid}, mark repeat = 15\\
color31, dashed, very thick,mark=*, mark options={solid} , mark repeat = 15\\
color32, dashed, very thick ,mark=*, mark options={solid}, mark repeat = 15\\
color33, dashed, very thick,mark=*, mark options={solid} , mark repeat = 15\\
color11, very thick \\
color12, very thick \\
color13, very thick \\
color21, very thick \\
color22, very thick \\
color23, very thick \\
color31, very thick \\
color32, very thick \\ 
color33, very thick \\
}

\pgfplotscreateplotcyclelist{mycolorlistWFXcell}{%
only marks,solid, mark options={solid,fill=color11,fill opacity=1,mark size=4pt},mark=square*,color11\\
only marks,solid, mark options={solid,fill=color22,fill opacity=1,mark size=4pt},mark=square*,color22\\
only marks,solid, mark options={solid,fill=color33,fill opacity=1,mark size=4pt},mark=square*,color33\\
color11, dashed, very thick,mark=*, mark options={solid}, mark repeat = 15\\
color22, dashed, very thick,mark=*, mark options={solid}, mark repeat = 15\\
color33, dashed, very thick,mark=*, mark options={solid}, mark repeat = 15\\
color11, very thick \\
color22, very thick \\
color33, very thick \\
}

\pgfplotscreateplotcyclelist{mycolorlistvol}{%
color11, dashed, very thick,mark=*, mark options={solid}, mark repeat = 15\\
color21, very thick \\
color33, very thick \\
}

\pgfplotscreateplotcyclelist{mycolorlist2}{%
only marks,solid, mark options={solid,fill=color11,fill opacity=1,mark size=4pt},mark=square*,color11\\
only marks,solid, mark options={solid,fill=color12,fill opacity=1,mark size=4pt},mark=square*,color12\\
only marks,solid, mark options={solid,fill=color13,fill opacity=1,mark size=4pt},mark=square*,color13\\
only marks,solid, mark options={solid,fill=color21,fill opacity=1,mark size=4pt},mark=square*,color21\\
only marks,solid, mark options={solid,fill=color22,fill opacity=1,mark size=4pt},mark=square*,color22\\
only marks,solid, mark options={solid,fill=color23,fill opacity=1,mark size=4pt},mark=square*,color23\\
only marks,solid, mark options={solid,fill=color31,fill opacity=1,mark size=4pt},mark=square*,color31\\
only marks,solid, mark options={solid,fill=color32,fill opacity=1,mark size=4pt},mark=square*,color32\\
only marks,solid, mark options={solid,fill=color33,fill opacity=1,mark size=4pt},mark=square*,color33\\
color11 \\
color11 \\
color11, dashed, very thick,mark=*, mark options={solid}, mark repeat = 15\\
color12, dashed, very thick ,mark=*, mark options={solid}, mark repeat = 15\\
color13, dashed, very thick ,mark=*, mark options={solid}, mark repeat = 15\\
color21, dashed, very thick ,mark=*, mark options={solid}, mark repeat = 15\\
color22, dashed, very thick ,mark=*, mark options={solid}, mark repeat = 15\\
color23, dashed, very thick ,mark=*, mark options={solid}, mark repeat = 15\\
color31, dashed, very thick,mark=*, mark options={solid} , mark repeat = 15\\
color32, dashed, very thick ,mark=*, mark options={solid}, mark repeat = 15\\
color33, dashed, very thick,mark=*, mark options={solid} , mark repeat = 15\\
color11, very thick \\
color12, very thick \\
color13, very thick \\
color21, very thick \\
color22, very thick \\
color23, very thick \\
color31, very thick \\
color32, very thick \\ 
color33, very thick \\
color11 \\
color11 \\
}

\pgfplotscreateplotcyclelist{mycolorlist123}{%
only marks,solid, mark options={solid,fill=color11,fill opacity=1,mark size=4pt},mark=square*,color11\\
only marks,solid, mark options={solid,fill=color22,fill opacity=1,mark size=4pt},mark=square*,color22\\
only marks,dashed, mark options={solid,fill=color33,fill opacity=1,mark size=4pt},mark=square*,color33\\
color11, very thick,dashed, mark options={solid}, mark repeat = 15\\
color22, very thick , mark options={solid}, mark repeat = 15\\
color33, very thick,dashed , mark options={solid}, mark repeat = 15\\
}

\pgfplotscreateplotcyclelist{mycolorlistSNE}{%
only marks,solid, mark options={solid,fill=color11,fill opacity=1,mark size=4pt},mark=square*,color11\\
only marks,solid, mark options={solid,fill=color22,fill opacity=1,mark size=4pt},mark=square*,color22\\
only marks,dashed, mark options={solid,fill=color33,fill opacity=1,mark size=4pt},mark=square*,color33\\
color11, very thick, dashed,  mark options={solid}, mark repeat = 15,mark=*\\
color22, very thick, dashed , mark options={solid}, mark repeat = 15, mark=*\\
color33, very thick,dashed , mark options={solid}, mark repeat = 15, mark=*\\
color11, very thick \\
color22, very thick \\
color33, very thick \\
}
\newcommand{\ba}{\boldsymbol{a}}
\newcommand{\bb}{\boldsymbol{b}}

\newcommand{\bh}{\boldsymbol{h}}

\newcommand{\bt}{\boldsymbol{t}}

\newcommand{\bA}{\boldsymbol{A}}

\newcommand{\bC}{\boldsymbol{C}}

\newcommand{\bF}{\boldsymbol{F}}

\newcommand{\bI}{\boldsymbol{I}}

\newcommand{\bP}{\boldsymbol{P}}
\newcommand{\bQ}{\boldsymbol{Q}}

\newcommand{\bW}{\boldsymbol{W}}
\newcommand{\bX}{\boldsymbol{X}}
\newcommand{\bx}{\boldsymbol{x}}

\newcommand{\by}{\boldsymbol{y}}

\newcommand{\tr}{\operatorname{tr}}
\newcommand{\cof}{\operatorname{Cof}}

\newcommand{\GL}{\text{GL}}
\newcommand{\SO}{\text{SO}}

\newcommand{\cB}{{\mathcal{B}}}

\newcommand{\cG}{{\mathcal{G}}}

\newcommand{\cI}{{\mathcal{I}}}

\newcommand{\cP}{{\mathcal{P}}}

\newcommand{\bbR}{{\mathbb{R}}}

\newcommand{\mathrelu}[1]{[#1]_+}   % Notation for "positive part" operation
\newcommand{\hadamard}{\ast}       % Notation for element-wise product of vectors

\definecolor{color11a}{RGB}{236, 188, 0}
\definecolor{color12a}{RGB}{50, 67, 121}
\definecolor{color13}{RGB}{222, 222, 222}
\definecolor{color21a}{RGB}{92, 134, 196}
\definecolor{color22a}{RGB}{249, 156, 0}
\definecolor{color23}{RGB}{222, 222, 222}
\definecolor{color31}{RGB}{222, 222, 222}
\definecolor{color32}{RGB}{222, 222, 222}
\definecolor{color33a}{RGB}{191, 60, 60}

\colorlet{color11}{color11a!90!color13}
\colorlet{color12}{color12a!90!color13}
\colorlet{color21}{color21a!90!color13}
\colorlet{color22}{color22a!90!color13}
\colorlet{color33}{color33a!90!color13}
\definecolor{colorCPSa}{RGB}{0, 157, 129}

\colorlet{2yellow}{color11a!90!color13}
\colorlet{2darkblue}{color12a!90!color13}
\colorlet{2lightblue}{color21a!90!color13}
\colorlet{2orange}{color22a!90!color13}
\colorlet{2red}{color33a!90!color13}
\definecolor{2grey}{RGB}{222, 222, 222}
\colorlet{2green}{colorCPSa!90!color13}

\usetikzlibrary{shapes}
\usetikzlibrary{arrows.meta} % for arrow size

\tikzset{>=latex} % for LaTeX arrow head
\tikzstyle{node}        =[thick, circle, draw=black, text = black, minimum size=25, inner sep=0.5, outer sep=0.6]
\tikzstyle{node icnn}   =[node, color=2orange!10!black, fill=color22!25]
\tikzstyle{node in_out} =[node, color=2lightblue!10!black, fill=color21!25]
\tikzstyle{node inv_pot}=[node, color=2red!10!black, fill=color33!25]
\tikzstyle{node layer}  =[node, rectangle, color=2red!10!black, fill=color33!25]

\tikzstyle{connect}=[thick, black, shorten <=1, shorten >=1, very thick] %,line cap=round
\tikzstyle{connect arrow}=[connect, densely dashed, -{Latex[length=4,width=3.5]}, very thick]

\tikzstyle{connect2}=[thick, black, shorten <=1, shorten >=1, very thick] %,line cap=round
\tikzstyle{connect arrow2}=[connect2, -{Latex[length=4,width=3.5]}, very thick]

\tikzstyle{connect arrow constraint}=[connect arrow2, 2red, very thick]
\tikzstyle{connect constraint}=[connect2, 2red, very thick] %,line cap=round

\newcommand{\layernode}[6][1]{
    \node[node 2] (#4) at (#2, #3) {};
    \ifnum#6=1 
        \draw[domain=-4:4, shift={(#2, #3-0.1)}, thick, scale=0.05, smooth, variable=\x, black] plot ({\x}, {ln(1+exp(1.5*\x))});
    \else
        \ifnum#6=2 
            \draw[domain=-4:4, shift={(#2, #3)}, thick, scale=0.05, smooth, variable=\x, black] plot ({\x}, {\x});
        \else
            \ifnum#6=3 
                \draw[domain=-4:4, shift={(#2, #3-0.1)}, thick, scale=0.05, smooth, variable=\x, black] plot ({\x}, {1.5*max(\x,0)});
            \else
                \draw[domain=-4:4, shift={(#2, #3)}, thick, scale=0.05, smooth, variable=\x, black] plot ({\x}, {3*tanh(\x)});
            \fi
        \fi
    \fi
    \
    \node[above left] at (#2, #3) {\small #5};
    \ifnum#1=1 
        \draw[connect arrow] (#2, #3+0.7) -- (#4);
    \fi
}

\def\Xone{$\boldsymbol{\cI}$}
\def\Xtwo{$\boldsymbol{t}$}
\def\Potental{\psi}
\def\Yout{$\Potental^{\text{NN}}$}

\newcommand{\norm}[1]{\left\lVert#1\right\rVert}

\definecolor{color11a}{RGB}{236, 188, 0}
\definecolor{color12a}{RGB}{50, 67, 121}
\definecolor{color13}{RGB}{222, 222, 222}
\definecolor{color21a}{RGB}{92, 134, 196}
\definecolor{color22a}{RGB}{249, 156, 0}
\definecolor{color33a}{RGB}{191, 60, 60}
\definecolor{colorCPSa}{RGB}{0, 157, 129}

\colorlet{2yellow}{color11a!90!color13}
\colorlet{2darkblue}{color12a!90!color13}
\colorlet{2lightblue}{color21a!90!color13}
\colorlet{2orange}{color22a!90!color13}
\colorlet{2red}{color33a!90!color13}
\colorlet{2grey}{color13}
\colorlet{2green}{colorCPSa!90!color13}

\def\markerrepeat{15}
\def\markerphase{1}
\pgfplotscreateplotcyclelist{colorlist_ROL_PE}{%
    ultra thick, dashed, mark = *, mark size = 2.5pt,2red,mark repeat = \markerrepeat, mark phase = \markerphase,mark options=solid\\
    ultra thick, dashed, mark = square*, mark size = 2.5pt,2orange,mark repeat = \markerrepeat, mark phase = \markerphase,mark options=solid\\
        ultra thick, dashed, mark = triangle*, mark size = 2.5pt,2darkblue,mark repeat = \markerrepeat, mark phase = \markerphase,mark options=solid\\
    2red, ultra thick \\
    2orange, ultra thick\\
            2darkblue, ultra thick \\
}

\pgfplotscreateplotcyclelist{typesfive}{
{2darkblue, ultra thick,mark=triangle*, mark repeat = 20},
{2orange,  ultra thick, mark=square*, mark repeat = 20},
{2red, mark=*, mark repeat = 20, ultra thick}
}

\pgfplotscreateplotcyclelist{StressArea}{
{2red, mark=*, mark repeat = 20, ultra thick},
{2darkblue, ultra thick,mark=triangle*, mark repeat = 20},
{2orange,  ultra thick, mark=square*, mark repeat = 20},
}

\newcommand{\predictionplot}[1]{
    \begin{tikzpicture}
    
        \pgfplotstableread{#1}
            \prediction
            
        \begin{axis}
        [cycle list name=colorlist_ROL_PE,
            xlabel=$F_{11}$,
            ylabel=$P_{ij}$,
            xmin=-1, xmax=1, % x scale
                            legend style={at={(0.5,-0.175)},anchor=north,font=\large},
                                            legend columns = 3,
                                            xtick={-1,0,1},
                                            xticklabels={0.5,1,1.5}
            ]
        
             \addplot table[x = lambda, y = data_11] from \prediction; %node[pos=0.0]{\hspace{-2em}$\large\boldsymbol{P_{##1}}$};
             \addplot table[x = lambda, y = data_33] from \prediction; %node[pos=0.0]{\hspace{-2em}$\large\boldsymbol{P_{##1}}$};
             \addplot table[x = lambda, y = data_21] from \prediction; %node[pos=0.0]{\hspace{-2em}$\large\boldsymbol{P_{##1}}$};
            
                       \addlegendentryexpanded{$P_{11}$},
           \addlegendentryexpanded{$P_{33}$},
           \addlegendentryexpanded{$P_{21}$},

            \addplot table[x = lambda, y = pred_11] from \prediction;
            \addplot table[x = lambda, y = pred_33] from \prediction;
            \addplot table[x = lambda, y = pred_21] from \prediction;

    \end{axis}
        
    \end{tikzpicture}
}

\newcommand*{\tvalues}{0.00,0.2,...,1.00}
\newcommand*{\extrapoints}{}

\newcommand{\MSEtplot}[1]{
    \begin{tikzpicture}
    
        \pgfplotstableread{#1}
            \meantlosses
    
        \begin{axis}
                [cycle list name=typesfive,
                xlabel=$t$,
                ylabel=MSE,
                xmin=-0.02, xmax=1.02, % x scale
                ymode=log,
                legend style={at={(0.5,-0.175)},anchor=north,font=\large},
                legend columns = 4
                ]
                
                  \addlegendimage{empty legend}
    \addlegendentryexpanded{pICNN type}
    
            \foreach \pICNNtype in {1,2,3} {
                \addplot[name path=A, draw=none, forget plot] table[y = Type_\pICNNtype_stds_upper] from \meantlosses ;
                \addplot[name path=B, draw=none, forget plot] table[y = Type_\pICNNtype_stds_lower] from \meantlosses ;
                \addplot+[fill opacity=0.15, forget plot] fill between[of=A and B];
    
                \addplot+[] table[y = Type_\pICNNtype_mean] from \meantlosses ;
                \addlegendentryexpanded{\pICNNtype}
            }

            \draw[gray, dashed] \foreach \x in \tvalues { (\x, 1e-10) -- (\x, 1e2)};

            \extrapoints
            
        \end{axis}
    \end{tikzpicture}
}

\newcommand{\StressAreaPlot}[1]{
    \begin{tikzpicture}
    
        \pgfplotstableread{#1}
            \meantlosses
    
        \begin{axis}
                [cycle list name=StressArea,
                xlabel=$F_{11}$,
                ylabel=$P_{ij}$,
                xmin=0.5, xmax=1.5, % x scale
                legend style={at={(0.5,-0.175)},anchor=north,font=\large},
                legend columns = 3
                ]

                                \addplot[dashed,2red, thick,mark=*,mark repeat = 20, mark options={solid},mark size = 2.5pt] table[y = P11] from \meantlosses ;
                \addlegendentry{$P_{11}$}
                
                                \addplot[dashed,2orange, thick,mark=square*,mark repeat = 20, mark options={solid},mark size = 2.5pt] table[y = P33] from \meantlosses ;
                \addlegendentry{$P_{33}$}

                \addplot[dashed,2darkblue, thick,mark=triangle*,mark repeat = 20, mark options={solid},mark size = 2.5pt] table[y = P21] from \meantlosses ;
                \addlegendentry{$P_{21}$}

                             \addplot[name path=A, 2red, thick] table[y = P11_min] from \meantlosses ;
                \addplot[name path=B,2red, thick] table[y = P11_max] from \meantlosses ;
                \addplot[fill opacity=0.15,2red, thick] fill between[of=A and B];

                                             \addplot[name path=A, 2orange, thick] table[y = P33_min] from \meantlosses ;
                \addplot[name path=B,2orange, thick] table[y = P33_max] from \meantlosses ;
                \addplot[fill opacity=0.15,2orange, thick] fill between[of=A and B];

                                                             \addplot[name path=A, 2darkblue, thick] table[y = P21_min] from \meantlosses ;
                \addplot[name path=B,2darkblue,thick] table[y = P21_max] from \meantlosses ;
                \addplot[fill opacity=0.15,2darkblue,thick] fill between[of=A and B];

        \end{axis}
    \end{tikzpicture}
}

\title{Parametrised polyconvex hyperelasticity with\\physics-augmented neural networks\vspace{1ex}}

\author[*]{Dominik~K.~Klein}
\author[ ]{Fabian~J.~Roth}
\author[ ]{\\Iman~Valizadeh}
\author[ ]{Oliver~Weeger}

\affil[ ]{\footnotesize Cyber-Physical Simulation, 
Department of Mechanical Engineering,\protect\\Technical University of Darmstadt, 64293 Darmstadt, Germany}
\affil[*]{\footnotesize Corresponding author, email: klein@cps.tu-darmstadt.de}

\date{June 15, 2023}

\begin{document}

\maketitle

\par\noindent\rule{\textwidth}{0.4pt}
\begin{abstract}
\noindent
In the present work, neural networks are applied to formulate parametrised hyperelastic constitutive models. The models fulfill all common mechanical conditions of hyperelasticity by construction.
In particular, partially input-convex neural network (pICNN) architectures are applied based on feed-forward neural networks. Receiving two different sets of input arguments, pICNNs are convex in one of them, while for the other, they represent arbitrary relationships which are not necessarily convex. In this way, the model can fulfill convexity conditions stemming from mechanical considerations without being too restrictive on the functional relationship in additional parameters, which may not necessarily be convex. Two different models are introduced, where one can represent arbitrary functional relationships in the additional parameters, while the other is monotonic in the additional parameters.
As a first proof of concept, the model is calibrated to data generated with two differently parametrised analytical potentials, whereby three different pICNN architectures are investigated. In all cases, the proposed model shows excellent performance.  \end{abstract}
\vspace*{2ex}
\par\noindent\rule{\textwidth}{0.4pt}

\section{Introduction}

Convexity is a convenient property of mathematical functions in many applications. However, it also constraints the function space a model can represent. While for some applications, this constraint is well motivated, it is too restrictive for other use cases. Moreover, there are applications where a function can be motivated to be convex in some of its arguments, while it should not necessarily be convex in the other arguments.
The latter is usually the case for hyperelastic material models with parametric dependencies, such as process parameters in 3D printing which influence material properties \cite{valizadeh}, or microstructured materials with a parametrised geometry \cite{fernandez}. 
In the framework of hyperelasticity, the polyconvexity condition introduced by \textcite{ball1976} requires the associated energy potentials to be convex functions in several strain measures. 
However, there is generally no mechanical motivation for a hyperelastic potential to be convex in additional parameters on which it might depend. To reflect this, a modeling framework for parametrised polyconvex hyperelasticity should provide potentials which are convex in the arguments of the polyconvexity condition and can represent more general functional relationships in the additional parameters.
Finally, in some cases, further conditions such as monotonicity of the hyperelastic potential in some parameters can be motivated by physical considerations \cite{valizadeh}.

\medskip

In constitutive modeling, neural networks (NNs) can be applied to represent hyperelastic potentials. These highly flexible models are usually formulated to fulfill mechanical conditions relevant to hyperelasticity. Such models are precious in fields where highly flexible yet physically sensible models are required, such as the simulation of microstructured materials \cite{gaertner,kalina,kumar}. Furthermore, including mechanical conditions improves the model generalization \cite{klein2022b}, allowing for model calibrations with sparse data usually available from real-world experiments \cite{linka2023}.
For the construction of polyconvex potentials, several approaches exist \cite{klein2022a,tac2022, chen}, where the most noteworthy approaches are based on input-convex neural networks (ICNNs). Proposed by \textcite{amos}, this special network architecture has not only been succesfully applied in the framework of polyconvexity, but is also very attractive in, e.g., other physical applications which require convexity \cite{huang} and convex optimization \cite{calafiore}.
Besides this particular choice of network architecture, using invariants as strain measures ensures fulfillment of several mechanical conditions at once, e.g., objectivity and material symmetry. This is well-known from analytical constitutive modeling \cite{schroeder, ebbing} and also commonly applied in NN-based models \cite{kalina, klein2022b, linka, tac}. 
Finally, by embedding the NN-potential into a larger modeling framework, i.e., adding additional analytical terms, all common constitutive conditions of hyperelasticity can be fulfilled by construction, which was at first introduced for compressible material behavior by \textcite{linden}. Therein, models that fulfill all mechanical conditions by construction are denoted as \textit{physics-augmented neural networks} (PANNs).

In the literature, also parametrised models were proposed for different applications, both in the analytical \cite{valizadeh, valizadeh2022, wu} and in the NN context \cite{baldi, shojaee}, in particular also including hyperelastic NN material models \cite{linka2021,fernandez}. 
However, to the best of the authors' knowledge, none of the existing parametrised hyperelastic models based on NNs fulfill all constitutive conditions at the same time. In particular, no model fulfills the polyconvexity condition, which is of great importance when applying hyperelastic constitutive models in numerical applications such as the finite element method, as it leads to a stable numerical behavior \cite{ebbing}.

\medskip

To conclude, while parametrised and polyconvex models are well-established in the framework of NN-based constitutive modeling, the link between both still needs to be made.
In the present work, this is done by applying partially-input convex neural networks (pICNNs) as proposed by \textcite{amos}. Receiving two sets of input arguments, pICNNs are convex in one while representing arbitrary relationships for the other. 
With the model proposed in this work being an extension of \textcite{linden}, all common constitutive conditions of hyperelasticity are fulfilled by construction. In particular, the model fulfills several mechanical conditions by using polyconvex strain invariants as inputs, while the pICNN preserves the polyconvexity of the invariants. Furthermore, growth and normalisation terms ensure a physically sensible stress behavior of the model. Two cases are considered, one with an arbitrary functional relationship in the additional parameters and the other being monotonic in the additional parameters. To formulate the functional relationships, three different pICNN architectures with different complexities are applied.

\medskip

The outline of the manuscript is as follows. In \cref{sec:1d}, the convexity of function compositions is discussed. In \cref{sec:basics} the fundamentals of parametrised hyperelasticity are briefly introduced, which are then applied to the proposed PANN model in \cref{sec:model}. The applicability of the parametric architectures is demonstrated by calibrating it to data generated with two differently parametrised analytical potentials in \cref{sec:example}, followed by the conclusion in \cref{sec:conc}. 

\newpage

\noindent\textbf{Notation.} Throughout this work, scalars, vectors and second order tensors are indicated by $a$, $\ba$ and $\bA$, respectively. The second order identity tensor is denoted as $\bI$. Transpose and inverse are denoted as $\bA^T$ and $\bA^{-1}$, respectively. Furthermore, trace, determinant and cofactor are denoted by $\tr\bA$, $\det\bA$ and $\cof \bA:=\det(\bA)\bA^{-T}$.
The set of invertible second order tensors with positive determinant is denoted by $\GL^+(3):=\left\{\bX\in\bbR^{3\times 3}\,|\, \det \bX > 0\right\}$ and the special orthogonal group in $\bbR^3$ by $\SO(3):=\left\{\bX\in\bbR^{3 \times 3}\,|\,\bX^T\bX=\bI,\,\allowbreak \det\bX=1\right\}$.
For the function composition $f(g(x))$ the compact notation $(f\circ g) (x)$ is applied.
The Softplus, Sigmoid, and ReLu functions are denoted by $s(x)=\ln(1+e^x)$, $sm(x)=\frac{1}{1+e^{-x}}$, and $\mathrelu{x} = \max(x,0)$, respectively.
The element-wise product between vectors is denoted as $\ast$.%, while $\mathrelu{x} = \max(x,0)$ denotes the ReLu-function.

\section{Convexity of function compositions}\label{sec:1d}

\begin{figure}[t!]
    \centering
    \begin{subfigure}[b]{0.45\textwidth}
        \centering
        \resizebox{\textwidth}{!}{
        \begin{tikzpicture}
\begin{axis}[
        xlabel=$x$,
        ylabel=$y$,
        xmin=-5, xmax=5, % x scale
        ymin=-1.5, ymax=5.5, % y scale
        domain=-5:5,   % added, key improvements
        legend cell align={left},
        legend style={at={(0.5,-0.175)},anchor=north,font=\large},
        legend columns = 4
        ]

    \addplot [2darkblue, ultra thick, mark=triangle*, mark repeat = 2, mark size = 2.5pt]    {ln(1+exp(x))};
  %  \addlegendentry{$g_1(x)=s(x)$}
    
            \addplot [2red,ultra thick, mark=*,mark repeat = 2, mark size=2.5pt]    {ln(1+exp(-x))}; 
  %  \addlegendentry{$h(x)=\frac{1}{5}x^2-1$}

    \addplot [2orange, ultra thick, mark=square*,mark repeat = 2, mark size = 2.5pt]   {0.2*x^2-1};
   % \addlegendentry{$g_2(x)=s(-x)$}

%    \node[left,text=2red] at (-5.1,0) {$\large\boldsymbol{g_1(x)}$};   
    
 %   \node[left,text=2darkblue] at (-5.1,5.01) {$\large\boldsymbol{g_2(x)}$};  
        
  %  \node[left,text=2green] at (-5.1,4) {$\large\boldsymbol{h(x)}$};  
    
%        \addlegendentryexpanded{$g_1(x)$}
%    \addlegendentryexpanded{$g_2(x)$}
%    \addlegendentryexpanded{$h(x)$}

                \addlegendentryexpanded{$ g_1(x)$}
        \addlegendentryexpanded{$g_2(x)$}
        \addlegendentryexpanded{$h(x)$}

\end{axis}
\end{tikzpicture}
        }
        \caption{Convex functions}
    \end{subfigure}
    \begin{subfigure}[b]{0.45\textwidth}   
        \centering 
        \resizebox{\textwidth}{!}{
        \begin{tikzpicture}
\begin{axis}[
        xlabel=$x$,
        ylabel=$y$,
        xmin=-5, xmax=5, % x scale
        ymin=-0.3, ymax=4.3, % y scale
        domain=-5:5,   % added, key improvements
        legend style={at={(0.5,-0.175)},anchor=north,font=\large},
        legend columns = 4]

    \addplot [2darkblue, ultra thick, mark=triangle*, mark repeat = 2, mark size = 2.5pt]    {ln(1+exp(0.2*x^2-1))};
 %   \addlegendentry{$f_1(x)=(g_1\circ h)(x)$}
    \addplot [2red, ultra thick, mark=*, mark repeat=2, mark size = 2.5pt]    {ln(1+exp(-(0.2*x^2-1))};
 %   \addlegendentry{$f_2(x)=(g_2\circ h)(x)$}

  %  \node[left,text=2darkblue] at (-5.1,0) {$\large\boldsymbol{f_2(x)}$}; 
    
  %      \node[left,text=2red] at (-5.1,4) {$\large\boldsymbol{f_1(x)}$}; 
    %    \addlegendentryexpanded{$\textcolor{black}{{f_1(x)}}$}
    %    \addlegendentryexpanded{$\textcolor{black}{{f_2(x)}}$}

                \addlegendentryexpanded{$(g_1 \circ h)(x)$}
        \addlegendentryexpanded{$(g_2 \circ h)(x)$}

\end{axis}
\end{tikzpicture}
        }    
        \caption{Compositions of convex functions}
    \end{subfigure}
    \caption{Compositions of univariate convex functions. $h(x)=0.2\,x^2-1$, $g_1(x)=s(x)$, $g_2(x)=s(-x)$.%, where $s(x)$ denotes the Softplus function. 
    Note that $g_1(x)$ is convex and non-decreasing, thus the composite function $(g_1\circ h)(x)$ is convex. $g_2(x)$ is convex but decreasing, and the composite function $(g_2\circ h)(x)$ is not convex.}
    \label{fig:1d}
\end{figure}
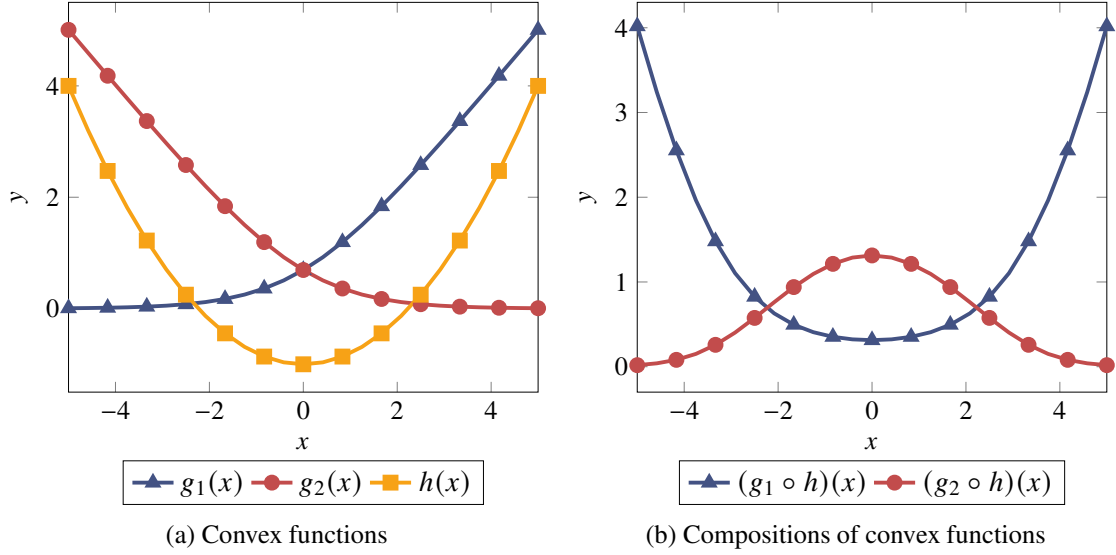

To lay the foundational intuition for constructing convex neural networks,  we first consider the univariate function
\begin{equation}\label{eq:1d}
    f:\bbR\rightarrow \bbR,\quad x \mapsto f(x):=(g\circ h)(x)\,,
\end{equation}
where $f$ is composed of two functions $g,h:\bbR\rightarrow\bbR$. Given that all of these functions are twice continuously differentiable, convexity of $f$ in $x$ is equivalent to the non-negativity of the second derivative
\begin{equation} \label{eq:convexity}
    f''(x)=(g''\circ h)(x)\,h'(x)^2 + (g'\circ h)(x) \,h'' (x) \geq 0\,.
\end{equation}
A sufficient, albeit not necessary condition for this is that the function $h$ is convex ($h'' \geq 0$), while the function $g$ is convex and non-decreasing ($g'\geq 0,\, g'' \geq 0$). Conversely, if a function acting on a convex function does not fulfill these conditions, the resulting function is not necessarily convex, see \cref{fig:1d} for an example. 
The recursive application of \cref{eq:convexity} yields conditions for arbitrary many function compositions. The innermost function, here $h$, must only be convex, while every following function must be convex and non-decreasing to preserve convexity.

The generalization to compositions of multivariate functions is also straightforward. For this, we consider the function
\begin{equation}
    f:\bbR^m\rightarrow \bbR,\quad \bx \mapsto f(\bx):=(g\circ \bh)(\bx)\,,
\end{equation}
with $\bh\colon\bbR^m\rightarrow\bbR^n$ and $g\colon\bbR^n\rightarrow\bbR$. Given that all of these functions are twice continuously differentiable, convexity of $f$ in $\bx$ is equivalent to the positive semi-definiteness of its Hessian. Similar reasoning as above leads to the sufficient condition that $\bh$ must be component-wise convex, while $g$ must be convex and non-decreasing, see \cite{klein2022a} for an explicit proof. Again, the recursive application of this yields conditions for arbitrary many function compositions. Here, the innermost function must be component-wise convex, while every following function must be component-wise convex and non-decreasing to preserve convexity.

\medskip

In the same manner, the composite function $f$, cf.~\cref{eq:1d}, is monotonically increasing (or non-decreasing) when its first derivative
\begin{equation}
        f'(x)=(g'\circ h)(x)\,h'(x) \geq 0
\end{equation}
is non-negative, which is fulfilled when both $g$ and $h$ are non-decreasing functions ($g'\geq 0,\,h'\geq 0$). The recursive application of this yields again conditions for arbitrary many function compositions. When all functions within a composite function are non-decreasing, the overall function is non-decreasing, see $(g_1\circ h)(x)$ for $x\geq 0$ in \cref{fig:1d} for an example. 
In this case, the generalization to compositions of vector-valued functions leads to the condition that all functions must be component-wise non-decreasing.

These basic ideas will be applied in both the mechanical requirements of the proposed model, cf.~\cref{sec:invs}, and in the construction of suitable network architectures, cf.~\cref{sec:architectures}.

\section{Parametrised hyperelastic constitutive modeling}
\label{sec:basics}

\subsection{Constitutive requirements for parametrised hyperelasticity}

The mechanical conditions of hyperelasticity are now briefly discussed. For a detailed introduction, the reader is referred to \cite{holzapfel,ebbing}.
The parametrised hyperelastic potential
\begin{equation}
    \psi\colon\GL^+(3)\times \bbR^n\rightarrow\bbR, \qquad \left(\bF;\,\bt\right)\mapsto \psi \left(\bF;\,\bt\right)
\end{equation}
corresponds to the strain energy density stored in the body $\cB\subset\bbR^3$ due to the deformation $\boldsymbol{\varphi}\colon\cB\rightarrow\bbR^3$. It depends on the deformation gradient $\bF=D\boldsymbol{\varphi}$ and the parameter vector $\bt\in\bbR^n$. With the stress being defined as the gradient field
\begin{equation}\label{eq:stress}
    \bP=\frac{\partial \psi\left(\bF;\,\bt\right)}{\partial \bF}\,,
\end{equation}
the \textbf{(i) second law of thermodynamics} is fulfilled by construction. 
The principle of \textbf{(ii) objectivity} states that a model should be independent on the choice of observer, which is formalized as
\begin{equation}
    \psi(\bQ\bF;\,\bt)=\psi(\bF;\,\bt)\qquad\forall\bF\in\GL^+(3),\,\bQ\in\SO(3),\,\bt\in\bbR^n\,.
\end{equation}
Also, the model should reflect the materials underlying (an-)isotropy, which corresponds to the \textbf{(iii) material symmetry condition}
\begin{equation}\label{eq:mat_sym}
\psi(\bF\bQ^T;\,\bt)=\psi(\bF;\,\bt)
\qquad\forall\bF \in\GL^+(3),\,\bQ\in\cG\subseteq \SO(3),\,\bt\in\bbR^n\,,
\end{equation}
where $\cG$ denotes the symmetry group under consideration. The \textbf{(iv) balance of angular momentum} implies that
\begin{equation}
    \frac{\partial \psi \left(\bF;\,\bt\right) }{\partial \bF}\bF^T=\bF\frac{\partial \psi \left(\bF;\,\bt\right)}{\partial \bF^T}
    \qquad\forall\bF\in\GL^+(3),\,\bt\in\bbR^n\,.
\end{equation}
Furthermore, we consider \textbf{(v) polyconvex} potentials which allow for a representation
\begin{equation}\label{eq:pc}
    \psi \left(\bF;\,\bt\right)=\cP\left(\boldsymbol{\xi};\,\bt\right)\,\qquad \text{with }\; \boldsymbol{\xi}:=\left(\bF,\,\cof\bF,\,\det\bF\right)\,,
\end{equation}
where $\cP$ is a convex function in $\boldsymbol{\xi}$. Note that polyconvexity does not restrict the potential's functional dependency on $\bt$. While the notion of polyconvexity stems from a rather theoretical context, it is also of practical relevance as it is the most straightforward way of fulfilling the ellipticity condition
\begin{equation}
    \left(\ba\otimes\bb\right)\colon\frac{\partial^2  \psi \left(\bF;\,\bt\right)}{\partial \bF\partial\bF}\colon \left(\ba\otimes\bb\right)\geq 0 \qquad \forall \ba,\bb\in\bbR^3\,.
\end{equation}
Also known as material stability, this condition leads to a favorable behavior in numerical applications. Finally, a physically sensible stress behavior requires fulfillment of the \textbf{(vi) growth condition}
\begin{equation}\label{eq:growth}
     \psi\rightarrow \infty \quad \text{as}\qquad \det\bF \rightarrow 0^+ \,,
\end{equation}
as well as a stress-free reference configuration $\bF=\bI$, also referred to as \textbf{(vii) normalisation}
\begin{equation}\label{eq:norm}
    \bP(\bI;\,\bt)=\boldsymbol{0}\qquad\forall\,\bt\in\bbR^n\,.
\end{equation}
In the most general case, no mechanical condition restricts the functional dependency of the potential $ \psi \left(\bF;\,\bt\right)$ in the parameters $\bt$. However, for \emph{some} applications, it may be well motivated to assume that the potential is a monotonically increasing function in the parameters. This \textbf{(viii) monotonicity condition} is formalized as 
\begin{equation}\label{eq:W_monoton}
    \frac{\partial  \psi \left(\bF;\,\bt\right)}{\partial t_i}\geq 0 \qquad\forall\,i\in \mathbb{N}_{\leq n},\,\bF\in\GL^+(3),\,\bt\in\bbR^n\,.
\end{equation}
Note that this does not imply monotonicity of the components of $ \bP(\bF;\,\bt)$ in $\bt$, which would mean that every component of the mixed second derivative
\begin{equation}\label{eq:P_monoton}
    \frac{\partial \bP(\bF;\,\bt)}{\partial \bt}=\frac{\partial^2 \psi\left(\bF;\,\bt\right)}{\partial \bF\partial \bt}
\end{equation}
would have to be non-negative. However, formulations which fulfill Eq.~\eqref{eq:P_monoton} could easily become too restrictive. E.g., they might lead to potentials which are convex in $\bF$ alone instead of the extended set of arguments of the polyconvexity condition, cf.~Eq.~\eqref{eq:pc}. However, convexity of the potential in $\bF$ is not compatible with a physically sensible material behavior \cite{gao}. Thus, the monotonicity condition Eq.~\eqref{eq:W_monoton} is applied throughout this work. 

Note that additional conditions on a physically sensible behavior of the hyperelastic potential can be formulated, e.g., the energy normalisation $\psi(\bI;\,\bt)=0\,\forall\,\bt\in\bbR^n$ \cite{linden}. However, throughout this work, we focus on the representation of the stress, meaning the gradient of the potential. Still, most conditions presented in this section are formulated in the hyperelastic potential, mainly for a convenient, brief notation.

\subsection{Invariant-based modeling}\label{sec:invs}
 
By formulating the potential $\psi$ in terms of invariants of the right Cauchy-Green tensor $\bC=\bF^T\bF$, conditions \textbf{(ii--iv)} can be fulfilled. Throughout this work, isotropic material behavior is assumed, i.e., $\cG=\SO(3)$ in \cref{eq:mat_sym}. In this case, three polyconvex invariants
 \begin{equation}\label{eq:invs}
     I_1=\tr\bC\,,\qquad I_2=\tr(\cof\bC)\,,\qquad I_3=\det\bC\,,
 \end{equation}
are considered. Then, the potential can be reformulated as\footnote{Note that $\psi \left(\bF;\,\bt\right)$ and $\psi\left(\boldsymbol{\cI};\,\bt\right)$ are different functions, but in the interest of readability, the same symbols are used.}  
\begin{equation} \label{eq:pot_invs}
    \psi:\bbR^m\times\bbR^n\rightarrow\bbR,\qquad \left(\boldsymbol{\cI};\,\bt\right)\mapsto\psi\left(\boldsymbol{\cI};\,\bt\right)\,,
\end{equation}
with 
\begin{equation}
    \boldsymbol{\cI}=\left(I_1,\,I_2,\,I_3,\,I_3^*\right)\in\bbR^4\,,\qquad     I_3^*=-\sqrt{I_3}\,,
\end{equation}
where the additional polyconvex invariant $I_3^*$ is important for the model to represent negative stress values, cf.~\cite{klein2022a}.
The invariants are non-linear functions in the arguments of the polyconvexity condition, cf.~\cref{eq:pc}. Thus, following \cref{sec:1d}, the potential $\psi$ must be \emph{convex} and component-wise \emph{non-decreasing} in $\boldsymbol{\cI}$ to preserve the polyconvexity of the invariants. By this, the overall potential fulfills the \textbf{(v) polyconvexity condition}. 
Note that this general form of the potential does not yet fulfill conditions \textbf{(vi--vii)}, which ensure a physically sensible stress behavior of the model.

\medskip

In the analytical case, an explicit choice of functional relationship for the hyperelastic potential has to be made, which fulfills all above introduced conditions. One such choice is the Neo-Hookean model
\begin{equation}\label{eq:NH_param}
     \psi^{\text{nh}}(I_1,I_3; \,t) = \frac{\mu(t)}{2}\left(I_1 - 3 - 2\ln \sqrt{I_3}\right) + \frac{\lambda(t)}{2}\left(\sqrt{I_3}-1\right)^2\,.
 \end{equation}
Here, the Lamé parameters $\lambda(t),\,\mu(t)$ are parametrised in terms of $t\in\bbR$.
While some analytical models base their functional relationship on physical reasoning, such as the Hencky model \cite{hencky, neff2016}, most constitutive models are of heuristic nature.
Simply put, the fulfillment of the objectivity condition by the Neo-Hookean model  has a solid mechanical motivation, while its linear dependency on $I_1$ has not and is simply a man-made choice.
The following section discusses how such limitations can be circumvented by applying NNs as highly flexible functions.
\section{Parameterised, physics-augmented neural network model}\label{sec:model}

As discussed in the previous section, the formulation of parametrised polyconvex potentials requires functions that are convex and non-decreasing in several strain invariants. At the same time, the functional relationship in the additional parameters should be either a general one or monotonically increasing, respectively, cf.~Eq.~\eqref{eq:W_monoton}.
Instead of making an explicit choice for such a formulation, we represent it by a neural network (NN), which can generally represent arbitrary functions \cite{hornik}. 

\subsection{Physics-augmented model formulation}

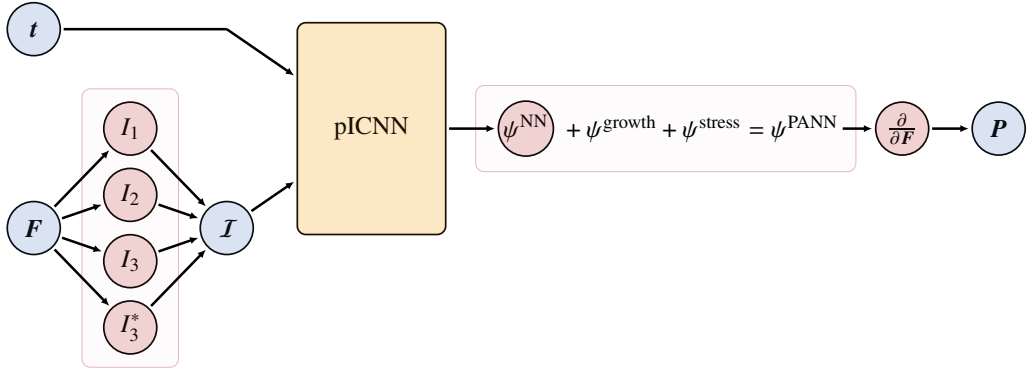
\begin{figure}
    \centering
    \resizebox{0.831\textwidth}{!}{
        \begin{tikzpicture}[x=1.6cm,y=1.1cm]

    \tikzset{ % node styles, numbered for easy mapping with \nstyle
        node 1/.style={node in_out},
        node 2/.style={node inv_pot},
        node 3/.style={node icnn},
        node 4/.style={node layer}
    }
    
    \large
    
    % Input layer
    \def\Xoff{0}
    \def\Ysep{3}
    \node[node 1] (t) at (\Xoff,  0.5*\Ysep) {$\bt$};
    \node[node 1] (F) at (\Xoff, -0.5*\Ysep) {$\bF$};
    
    % Invariant layer
    \pgfmathparse{\Xoff+1}
    \xdef\Xoff{\pgfmathresult}
    \def\Yoff{-1.5}
    \def\Ysep{1}

    \draw[color33!40,fill=color33,fill opacity=0.02,rounded corners=4]
        (\Xoff-0.5,\Yoff-1.5*\Ysep-0.6) rectangle++ (1, 3*\Ysep+1.2);
    
    \node[node 2] (I1)  at (\Xoff, \Yoff + 1.5*\Ysep) {$I_1$};
    \node[node 2] (I2)  at (\Xoff, \Yoff + 0.5*\Ysep) {$I_2$};
    \node[node 2] (I3)  at (\Xoff, \Yoff - 0.5*\Ysep) {$I_3$};
    \node[node 2] (I3*) at (\Xoff, \Yoff - 1.5*\Ysep) {$I_3^*$};

    \draw[connect arrow2] (F) -- (I1);
    \draw[connect arrow2] (F) -- (I2);
    \draw[connect arrow2] (F) -- (I3);
    \draw[connect arrow2] (F) -- (I3*);

    % pICNN Input Layer
    \pgfmathparse{\Xoff+1}
    \xdef\Xoff{\pgfmathresult}
    \def\Ysep{3}
    
    %\node[node 1] (X2) at (\Xoff,  0.5*\Ysep) {\Xtwo};
    \node[node 1] (X1) at (\Xoff, -0.5*\Ysep) {\Xone};

    \draw[connect arrow2] (I1)  -- (X1);
    \draw[connect arrow2] (I2)  -- (X1);
    \draw[connect arrow2] (I3)  -- (X1);
    \draw[connect arrow2] (I3*) -- (X1);

    \draw[connect, shorten >=0] (t) -- (\Xoff,  0.5*\Ysep);

    % pICNN
    \pgfmathparse{\Xoff+1.5}
    \xdef\Xoff{\pgfmathresult}

    \node (pICNN) at (\Xoff,0) [node 3, rectangle, minimum width=70, minimum height=100, rounded corners=4] {pICNN};

    \draw[connect arrow2, shorten <=0] (\Xoff-1.5,  0.5*\Ysep) -- (pICNN);
    %\draw[connect arrow] (t) -- (pICNN);
    \draw[connect arrow2] (X1) -- (pICNN);

    % Potential layer
    \pgfmathparse{\Xoff+1.6}
    \xdef\Xoff{\pgfmathresult}
    \def\Border{1.3}
    \def\additionXoff{1.3}

    \draw[color33!40,fill=color33,fill opacity=0.02,rounded corners=4]
        (\Xoff-0.4*\Border, -0.5*\Border) rectangle (\Xoff + \additionXoff + 0.4*\Border + 0.7+0.9, 0.5*\Border);
    
    \node[node 2, outer sep=0.6] (W) at (\Xoff,0) {$\psi^{\text{NN}}$};
    \draw[connect arrow2]  (pICNN) -- (W);
    
    \node[thick, inner sep=0.5, outer sep=0.6] (W) at (\Xoff+\additionXoff+0.5, 0) {${}+\psi^{\text{growth}}+\psi^{\text{stress}}=\psi^{\text{PANN}}$};

    % Differentiation layer
    \pgfmathparse{\Xoff+3.3}
    \xdef\Xoff{\pgfmathresult}
    
    \node[node 2, outer sep=0.6] (d_dF) at (\Xoff+0.6,0) {$\frac{\partial}{\partial\bF}$};
    \draw[connect arrow2]  (W) -- (d_dF);
    
    % Output layer
    \pgfmathparse{\Xoff+1}
    \xdef\Xoff{\pgfmathresult}
    \node[node 1] (P) at (\Xoff+0.6, 0) {$\bP$};
    \draw[connect arrow2]  (d_dF) -- (P);
 
  \end{tikzpicture}
    }
    \caption{Illustration of the PANN-based constitutive model. The pICNN is convex and non-decreasing in the invariants $\boldsymbol{\cI}$ while representing arbitrary (or monotonically increasing) functional relationships in the additional parameters $\bt$.}
    \label{fig:model_architecture}
\end{figure}

To incorporate the constitutive requirements introduced above in \cref{sec:basics}, the NN is only a part of the overall PANN material model given by
\begin{equation}\label{eq:PANN}
    \psi^{\text{PANN}}(\boldsymbol{\cI};\,\bt)=\psi^{\text{NN}}(\boldsymbol{\cI};\,\bt)+\psi^\text{growth}(J)+\psi^\text{stress}(J;\,\bt)\,,
\end{equation}
which is an extension of the model proposed by \textcite{linden} with parametric dependencies. The overall flow and structure of the model are visualized in \cref{fig:model_architecture}. 

In \cref{eq:PANN}, $\psi^{\text{NN}}(\boldsymbol{\cI};\,\bt)$ denotes the pICNN, which is convex and non-decreasing in $\boldsymbol{\cI}$ and arbitrary (or monotonically increasing) in $\bt$. In \cref{sec:architectures}, different pICNN architectures are discussed. To this point, $\psi^{\text{NN}}$ is treated as a general, sufficiently smooth function. 
The remaining terms in \cref{eq:PANN} ensure a physically sensible stress behavior of the model. In particular, they ensure the growth and normalisation conditions, cf. Eqs.~(\ref{eq:growth},\ref{eq:norm}).
With the analytical growth term 
\begin{equation}
    \psi^{\text{growth}}(J):=\left(J+\frac{1}{J}-2\right)^2
\end{equation}
and the normalisation term introduced by \cite{linden}
\begin{equation}
    \psi^{\text{stress}}(J;\,\bt):=-\mathfrak{n}(\bt)\,J\,,
\end{equation}
the polyconvexity of the model is preserved, cf. \cite{linden} for a discussion. Here, 
\begin{equation}
    \mathfrak{n}(\bt):=2\left(\frac{\partial \psi^\text{NN}}{\partial I_1}(\bt) 
    + 2\frac{\partial \psi^\text{NN}}{\partial I_2}(\bt) + \frac{\partial \psi^\text{NN}}{\partial I_3}(\bt)  -\frac{\partial \psi^\text{NN}}{\partial I_3^*}  (\bt)
    \right)\Bigg\rvert_{\bF=\bI}\in\bbR
\end{equation}
is a weighted sum of derivatives of the pICNN potential with respect to the invariants for the undeformed state $\bF=\bI$.

In most applications, the stress, meaning the gradient of the potential, c.f.~\cref{eq:stress}, is of interest rather than the potential itself. Here, the gradient of the potential can be evaluated either by using automatic differentiation, or by calculating the derivatives of the NN potential in an explicit way, cf.~\cite{franke}.

\subsection{Partially input-convex neural network architectures}\label{sec:architectures}

Different pICNN architectures applicable to the model are now discussed, which are all based on FFNNs. From a formal point of view, FFNNs are multiple compositions of vector-valued functions \cite{aggarwal}. The components are referred to as nodes or neurons, and the function acting in each node is referred to as activation function. 
The simple structure and recursive definition of FFNNs make them a very natural choice for constructing convex functions.
In a nutshell, when the first layer is component-wise convex and every subsequent layer is component-wise convex and non-decreasing, the overall function is convex in its input, cf. \cref{sec:1d}. This can also be adapted to partially convex functions, as proposed by \textcite{amos}.

\medskip

\begin{definition}[\textbf{Partially input-convex neural networks (pICNNs)}]
The FFNN 
\begin{equation}
    \cP\colon \mathbb{R}^m\times\mathbb{R}^n\rightarrow \mathbb{R}\,,\qquad \left(\bx,\,\by\right) \mapsto \cP\left(\bx,\,\by\right)
\end{equation}
is called a pICNN, if $\cP$ is convex w.r.t. $\bx$.  
\end{definition} 

\medskip

In the following, three different pICNN architectures are described. The interrelation between the two inputs and the overall complexity gets gradually more pronounced from Type 1 to Type 3, with Type 3 being a slightly adapted version of the architecture proposed by \textcite{amos}. The more complex pICNN architectures can be reduced to the simpler ones by constraining a subset of their parameters to take on specific values. For explicit proofs of convexity, the reader is referred to \cite{klein2022a} and made aware of the fact that, when investigating convexity in $\bx$, the influence of the non-convex input $\by$ can be seen as an additional bias which does not influence convexity in $\bx$. In addition, an adapted version of the simplest pICNN architecture, which is monotonically increasing in $\by$, is discussed. In general, also the other two pICNN architectures could be adapted to be monotonically increasing in $\by$.

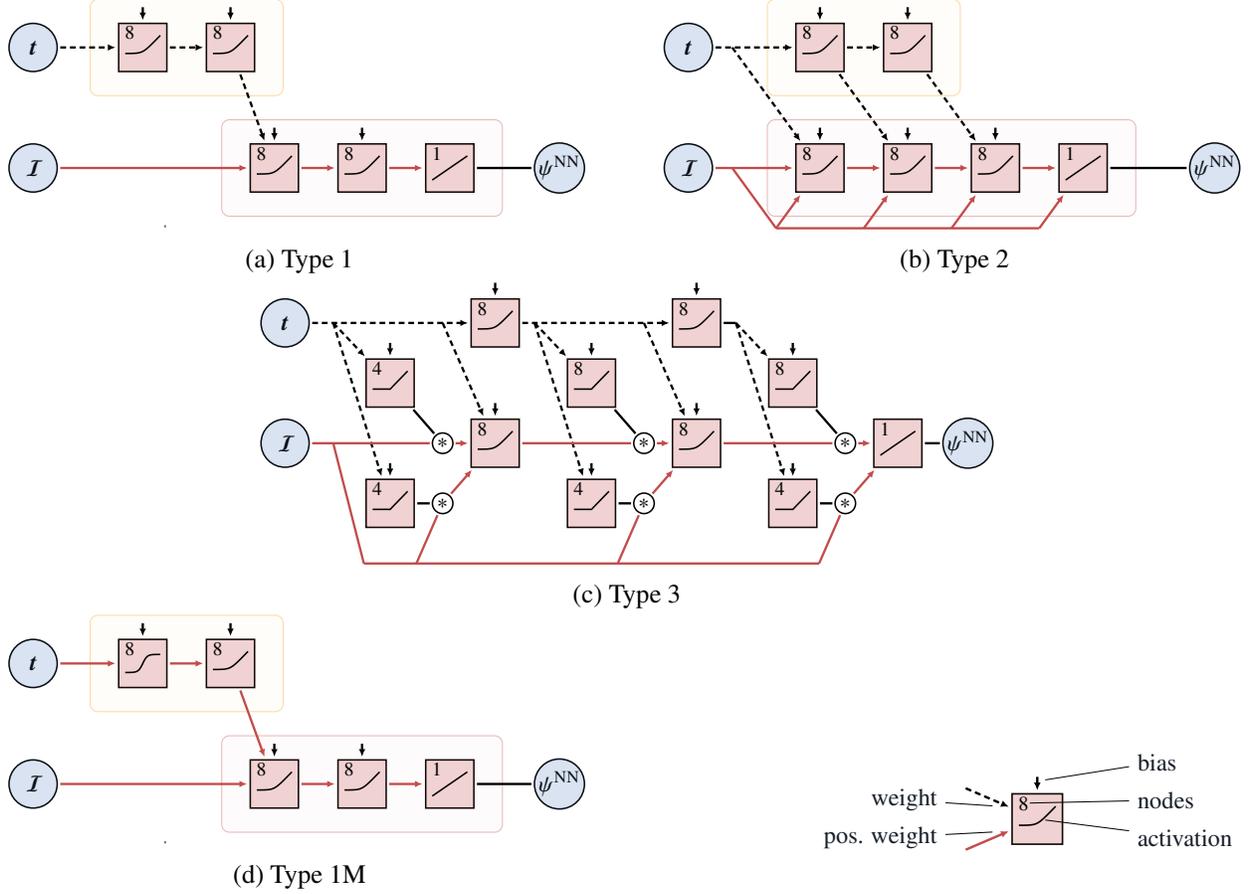
\begin{figure}
   \centering
    \begin{subfigure}[b]{0.475\textwidth}
        \centering
        \resizebox{\textwidth}{!}{
            %\tikzsetnextfilename{NN_figure}

\begin{tikzpicture}[x=1.6cm,y=1.1cm]
    \large
    
    \tikzset{ % node styles, numbered for easy mapping with \nstyle
    node 1/.style={node in_out},
    node 2/.style={node layer}
    }
    
    % rectangles
    \draw[2orange!40, fill=2orange, fill opacity=0.02, rounded corners=4]
    (2.15,2.2) rectangle++ (2.2,1.6);
    \draw[color33!40, fill=color33, fill opacity=0.02, rounded corners=4]
    (3.65,0.2) rectangle++ (3.2,1.6);

    % Input layer
    \node[node 1] (X1) at (1.5, 1) {\Xone};
    \node[node 1] (X2) at (1.5, 3) {\Xtwo};

    % Non-convex part
    \layernode{2.75}{3}{nc-1}{8}{1}
    \layernode{3.75}{3}{nc-2}{8}{1}

    % Convex part
    \layernode{4.25}{1}{c-1}{8}{1}
    \layernode{5.25}{1}{c-2}{8}{1}
    \layernode[0]{6.25}{1}{c-3}{1}{2}

    \draw[connect arrow] (X2) -- (nc-1);
    \draw[connect arrow] (nc-1) -- (nc-2);
    \draw[connect arrow] (nc-2) -- (c-1);
    %\node[left] at (4, 2) {\small $\pm$};
    
    \draw[connect arrow constraint] (X1) -- (c-1);
    \draw[connect arrow constraint] (c-1) -- (c-2);
    \draw[connect arrow constraint] (c-2) -- (c-3);

    % Output layer
    \node[node 1] (Y) at (7.5, 1) {\Yout};

    \draw[connect] (c-3) -- (Y);

    % Invisible passthrough layer for better alignment with Type C (Type 2, or whatever)
    \draw[connect] (3,0) -- (3,0);
    
    % labels
    % \node[align=center,color21!60!black] at (0.75,-3.5) {input};
    % \node[align=center,color21!60!black] at (10.75,-3.5) {output};
    % \node[align=center,color33!60!black] at (2.5,-3.5) {polyconvex invariants:\\[-0.2em]objectivity and\\[-0.2em]material symmetry};
    % \node[align=center,color22!60!black] at (5.25, -3.5) {ICNN:\\[-0.2em]flexibility};
    % \node[align=center,color33!60!black] at (8.625,-3.5) {potential:\\[-0.2em]coercivity and thermo-\\[-0.2em]dynamic consistency};
 
  \end{tikzpicture}
        }
        \caption{Type 1}
        \label{fig:pICNN architecture type A}
    \end{subfigure}
    \hfill
    \begin{subfigure}[b]{0.475\textwidth}   
        \centering 
        \resizebox{\textwidth}{!}{
            \begin{tikzpicture}[x=1.6cm,y=1.1cm]
    \large
    
    \tikzset{ % node styles, numbered for easy mapping with \nstyle
    node 1/.style={node in_out},
    node 2/.style={node layer}
    }
    
    % rectangles
    \draw[color22!40, fill=color22, fill opacity=0.02, rounded corners=4]
    (1.4,2.2) rectangle++ (2.2,1.6);
    \draw[color33!40, fill=color33, fill opacity=0.02, rounded corners=4]
    (1.4,0.2) rectangle++ (4.2,1.6);

    % Input layer
    \node[node 1] (X1) at (0.5, 1) {\Xone};
    \node[node 1] (X2) at (0.5, 3) {\Xtwo};

    % Non-convex part
    \layernode{2}{3}{nc-1}{8}{1}
    \layernode{3}{3}{nc-2}{8}{1}

    % Convex part
    \layernode{2}{1}{c-1}{8}{1}
    \layernode{3}{1}{c-2}{8}{1}
    \layernode{4}{1}{c-3}{8}{1}
    \layernode[0]{5}{1}{c-4}{1}{2}

    \draw[connect arrow] (X2) -- (nc-1);
    \draw[connect arrow, shorten <=0] (1,3) -- (c-1);
    \draw[connect arrow] (nc-1) -- (nc-2);
    \draw[connect arrow] (nc-1) -- (c-2);
    \draw[connect arrow] (nc-2) -- (c-3);
    
    \draw[connect arrow constraint] (X1) -- (c-1);
    \draw[connect arrow constraint] (c-1) -- (c-2);
    \draw[connect arrow constraint] (c-2) -- (c-3);
    \draw[connect arrow constraint] (c-3) -- (c-4);

    \draw[connect constraint, shorten <=0, shorten >=0] (1,1) -- (1.5,0);
    \draw[connect constraint, shorten <=0, shorten >=0] (1.5,0) -- (4.5,0);
    \draw[connect arrow constraint, shorten <=0] (1.5,0) -- (c-1);
    \draw[connect arrow constraint, shorten <=0] (2.5,0) -- (c-2);
    \draw[connect arrow constraint, shorten <=0] (3.5,0) -- (c-3);
    \draw[connect arrow constraint, shorten <=0] (4.5,0) -- (c-4);

    % Output layer
    \node[node 1] (Y) at (6.5, 1) {\Yout};

    \draw[connect] (c-4) -- (Y);
    
    % labels
    % \node[align=center,color21!60!black] at (0.75,-3.5) {input};
    % \node[align=center,color21!60!black] at (10.75,-3.5) {output};
    % \node[align=center,color33!60!black] at (2.5,-3.5) {polyconvex invariants:\\[-0.2em]objectivity and\\[-0.2em]material symmetry};
    % \node[align=center,color22!60!black] at (5.25, -3.5) {ICNN:\\[-0.2em]flexibility};
    % \node[align=center,color33!60!black] at (8.625,-3.5) {potential:\\[-0.2em]coercivity and thermo-\\[-0.2em]dynamic consistency};
 
  \end{tikzpicture}
        }    
        \caption{Type 2}
        \label{fig:pICNN architecture type C}
    \end{subfigure}
  %  \hfill
              \begin{subfigure}[b]{0.6045\textwidth}   
        \centering 
        \resizebox{\textwidth}{!}{
            \newcommand{\hadamardnode}[3]{
    %           #1: x-coordinate
    %           #2: y-coordinate
    %           #3: node-name
    \node[thick, circle, draw=black, text = black, minimum size=10, inner sep=1] (#3) at (#1, #2) {$*$};
}

\begin{tikzpicture}[x=1.6cm,y=1.1cm]

    \def\Xhad{0.6} % hadamard node separation

    \large
    
    \tikzset{ % node styles, numbered for easy mapping with \nstyle
    node 1/.style={node in_out},
    node 2/.style={node layer}
    }
    
    % rectangles
    % \draw[color22!40, fill=color22, fill opacity=0.02, rounded corners=4]
    % (1.4,2.2) rectangle++ (2.2,1.6);
    % \draw[color33!40, fill=color33, fill opacity=0.02, rounded corners=4]
    % (1.4,0.2) rectangle++ (4.2,1.6);

    % Input layer
    \node[node 1] (X1) at (0.5, 1) {\Xone};
    \node[node 1] (X2) at (0.5, 3) {\Xtwo};

    % Layer 1
    \def\Xzero{1.7}
    
    \layernode{\Xzero}{2}{pnc-11}{4}{3}
    \layernode{\Xzero}{0}{pnc-12}{4}{3}

    \hadamardnode{\Xzero+\Xhad}{1}{h-11}
    \hadamardnode{\Xzero+\Xhad}{0}{h-12}

    \layernode{\Xzero+2*\Xhad}{3}{nc-1}{8}{1}
    \layernode{\Xzero+2*\Xhad}{1}{c-1}{8}{1}

    \draw[connect arrow] (X2) -- (nc-1);
    \draw[connect arrow, shorten <=0] (\Xzero-0.65,3) -- (pnc-11);
    \draw[connect arrow, shorten <=0] (\Xzero-0.65,3) -- (pnc-12);

    \draw[connect] (pnc-11) -- (h-11);
    \draw[connect] (pnc-12) -- (h-12);

    \draw[connect constraint] (X1) -- (h-11);

    \draw[connect arrow constraint] (h-11) -- (c-1);
    \draw[connect arrow constraint] (h-12) -- (c-1);
    \draw[connect arrow, shorten <=0] (\Xzero+\Xhad, 3) -- (c-1);

    \draw[connect constraint, shorten <=0] (\Xzero+0.5*\Xhad,-1) -- (h-12);

    % Layer 2
    \def\Xzero{4}
    
    \layernode{\Xzero}{2}{pnc-21}{8}{3}
    \layernode{\Xzero}{0}{pnc-22}{4}{3}

    \hadamardnode{\Xzero+\Xhad}{1}{h-21}
    \hadamardnode{\Xzero+\Xhad}{0}{h-22}

    \layernode{\Xzero+2*\Xhad}{3}{nc-2}{8}{1}
    \layernode{\Xzero+2*\Xhad}{1}{c-2}{8}{1}

    \draw[connect arrow] (nc-1) -- (nc-2);
    \draw[connect arrow, shorten <=0] (\Xzero-0.65,3) -- (pnc-21);
    \draw[connect arrow, shorten <=0] (\Xzero-0.65,3) -- (pnc-22);

    \draw[connect] (pnc-21) -- (h-21);
    \draw[connect] (pnc-22) -- (h-22);

    \draw[connect constraint] (c-1) -- (h-21);

    \draw[connect arrow constraint] (h-21) -- (c-2);
    \draw[connect arrow constraint] (h-22) -- (c-2);
    \draw[connect arrow, shorten <=0] (\Xzero+\Xhad, 3) -- (c-2);

    \draw[connect constraint, shorten <=0] (\Xzero+0.5*\Xhad,-1) -- (h-22);

    % Layer 3
    \def\Xzero{6.3}
    
    \layernode{\Xzero}{2}{pnc-31}{8}{3}
    \layernode{\Xzero}{0}{pnc-32}{4}{3}

    \hadamardnode{\Xzero+\Xhad}{1}{h-31}
    \hadamardnode{\Xzero+\Xhad}{0}{h-32}

    %\layernode{\Xzero+2*\Xhad}{3}{nc-3}{8}{1}
    \layernode[0]{\Xzero+2*\Xhad}{1}{c-3}{1}{2}

    %\draw[connect arrow] (nc-2) -- (nc-3);
    \draw[connect, shorten >=0] (nc-2) -- (\Xzero-0.65,3);
    \draw[connect arrow, shorten <=0] (\Xzero-0.65,3) -- (pnc-31);
    \draw[connect arrow, shorten <=0] (\Xzero-0.65,3) -- (pnc-32);

    \draw[connect] (pnc-31) -- (h-31);
    \draw[connect] (pnc-32) -- (h-32);

    \draw[connect constraint] (c-2) -- (h-31);

    \draw[connect arrow constraint] (h-31) -- (c-3);
    \draw[connect arrow constraint] (h-32) -- (c-3);
    %\draw[connect arrow, shorten <=0] (\Xzero+\Xhad, 3) -- (c-3);

    \draw[connect constraint, shorten <=0] (\Xzero+0.5*\Xhad,-1) -- (h-32);
    
    % Passthrough connection
    \draw[connect constraint, shorten <=0, shorten >=0] (1.05,1) -- (1.4,-1);
    \draw[connect constraint, shorten <=0, shorten >=0] (1.4,-1) -- (\Xzero+0.5*\Xhad,-1);
    
    % Output layer
    \node[node 1] (Y) at (\Xzero+2, 1) {\Yout};

    \draw[connect] (c-3) -- (Y);
    
    % labels
    % \node[align=center,color21!60!black] at (0.75,-3.5) {input};
    % \node[align=center,color21!60!black] at (10.75,-3.5) {output};
    % \node[align=center,color33!60!black] at (2.5,-3.5) {polyconvex invariants:\\[-0.2em]objectivity and\\[-0.2em]material symmetry};
    % \node[align=center,color22!60!black] at (5.25, -3.5) {ICNN:\\[-0.2em]flexibility};
    % \node[align=center,color33!60!black] at (8.625,-3.5) {potential:\\[-0.2em]coercivity and thermo-\\[-0.2em]dynamic consistency};
 
  \end{tikzpicture}
        }
        \caption{Type 3}
        \label{fig:pICNN architecture type D}
    \end{subfigure}
                    \begin{subfigure}[b]{0.475\textwidth}
        \centering
        \resizebox{\textwidth}{!}{
            %\tikzsetnextfilename{NN_figure}

\begin{tikzpicture}[x=1.6cm,y=1.1cm]
    \large
    
    \tikzset{ % node styles, numbered for easy mapping with \nstyle
    node 1/.style={node in_out},
    node 2/.style={node layer}
    }
    
    % rectangles
    \draw[2orange!40, fill=2orange, fill opacity=0.02, rounded corners=4]
    (2.15,2.2) rectangle++ (2.2,1.6);
    \draw[color33!40, fill=color33, fill opacity=0.02, rounded corners=4]
    (3.65,0.2) rectangle++ (3.2,1.6);

    % Input layer
    \node[node 1] (X1) at (1.5, 1) {\Xone};
    \node[node 1] (X2) at (1.5, 3) {\Xtwo};

    % Non-convex part
    \layernode{2.75}{3}{nc-1}{8}{4}
    \layernode{3.75}{3}{nc-2}{8}{1}

    % Convex part
    \layernode{4.25}{1}{c-1}{8}{1}
    \layernode{5.25}{1}{c-2}{8}{1}
    \layernode[0]{6.25}{1}{c-3}{1}{2}

    \draw[connect arrow constraint] (X2) -- (nc-1);
    \draw[connect arrow constraint] (nc-1) -- (nc-2);
    \draw[connect arrow constraint] (nc-2) -- (c-1);
    %\node[left] at (4, 2) {\small $\pm$};
    
    \draw[connect arrow constraint] (X1) -- (c-1);
    \draw[connect arrow constraint] (c-1) -- (c-2);
    \draw[connect arrow constraint] (c-2) -- (c-3);

    % Output layer
    \node[node 1] (Y) at (7.5, 1) {\Yout};

    \draw[connect] (c-3) -- (Y);

    % Invisible passthrough layer for better alignment with Type C (Type 2, or whatever)
    \draw[connect] (3,0) -- (3,0);
    
    % labels
    % \node[align=center,color21!60!black] at (0.75,-3.5) {input};
    % \node[align=center,color21!60!black] at (10.75,-3.5) {output};
    % \node[align=center,color33!60!black] at (2.5,-3.5) {polyconvex invariants:\\[-0.2em]objectivity and\\[-0.2em]material symmetry};
    % \node[align=center,color22!60!black] at (5.25, -3.5) {ICNN:\\[-0.2em]flexibility};
    % \node[align=center,color33!60!black] at (8.625,-3.5) {potential:\\[-0.2em]coercivity and thermo-\\[-0.2em]dynamic consistency};
 
  \end{tikzpicture}
        }
        \caption{Type 1M}
        \label{fig:pICNN architecture type A}
    \end{subfigure}
        \hfill
                \begin{subfigure}[b]{0.3530\textwidth}   
        \centering 
        \resizebox{\textwidth}{!}{
            \begin{tikzpicture}[x=1.6cm,y=1.1cm]

    \tikzstyle{leggend text} = [color21!20!black]

    \large
    
    \tikzset{ % node styles, numbered for easy mapping with \nstyle
    node 1/.style={node in_out},
    node 2/.style={node layer}
    }
    
    \tikzstyle{lconnect}=[connect, thin, shorten <=0, shorten >=4] %,line cap=round

    % Non-convex part
    \layernode{0}{0}{nc-1}{8}{1}

    \draw[connect arrow] (-0.8, 0.5) -- (nc-1);
    \draw[connect arrow constraint] (-0.8, -0.5) -- (nc-1);

    \node[anchor=west, leggend text] (bias) at (1,0.9) {bias};
    \draw[lconnect] (bias) -- (0, 0.6);

    \node[anchor=west, leggend text] (nodes) at (1, 0.3) {nodes};
    \draw[lconnect] (nodes) -- (-0.17, 0.25);

    \node[anchor=west, leggend text] (activation) at (1,-0.3) {activation};
    \draw[lconnect] (activation) -- (0, 0);

    \node[anchor=east, leggend text] (weight) at (-1, 0.3) {weight};
    \draw[lconnect] (weight) -- (-0.4, 0.2);

    \node[anchor=east, leggend text] (weight constraint) at (-1,-0.3) {pos. weight};
    \draw[lconnect] (weight constraint) -- (-0.4, -0.2);
    
  \end{tikzpicture}
        }
        %\caption{Legend}
        \label{fig:pICNN architecture legend} 
    \end{subfigure}
    \caption{Different pICNN architectures for the representation of the neural network potential $\psi^{\text{NN}}$. For Type 1-3, the NN is convex and non-decreasing in $\boldsymbol{\cI}$, and can take arbitrary functional relationships in $\boldsymbol{t}$. In addition, for Type 1M, the NN is monotonically increasing in $\bt$.
    }
    \label{fig:pICNN architectures}
\end{figure}

Note that for representing a parametrised polyconvex potential, the pICNN must be convex and \emph{non-decreasing} in $\bx$, as discussed in \cref{sec:invs}. This requires some adaptions to the general pICNN architectures. The adaptions are discussed after introducing the general architectures, and the adapted architectures are visualized in \cref{fig:pICNN architectures} for one specific choice of nodes and layers.

\medskip

\begin{prop}[\textbf{pICNN -- Type 1}] The pICNN with input $\bx=:\bx_0,\,\by=:\by_0$, output $\cP\left(\bx,\,\by\right):=\bx_{H_x+1}\in\bbR$ and $H_x,H_y$ hidden layers
\begin{equation}
    \begin{aligned}
        \by_{h+1} &= \sigma_h        \left(  \bW_h^{[yy]}\,\by_h + \bb_h^{[y]}\right)  &&\in\mathbb{R}^{\mathrlap{n_h}\phantom{m_h}},\quad h=0,\dotsc,H_y\,,
        \\
        \bx_{\mathrlap{1}\phantom{h+1}} &= \tilde{\sigma}_{\mathrlap{0}\phantom{h}}  \left(  \bW_0^{[xx]}\,\bx_0 + \bb_0^{[x]} + \bW^{[xy]}\,\by_{H_y+1} \right)  &&\in\mathbb{R}^{\mathrlap{m_0}\phantom{m_h}}, 
        \\
        \bx_{h+1} &= \tilde{\sigma}_h \left(  \bW_h^{[xx]}\,\bx_h + \bb_h^{[x]} \right)  &&\in\mathbb{R}^{m_h},\quad h=1,\dotsc,H_x
    \end{aligned}
\end{equation}
is convex in $\bx$ given that the weights $\bW_h^{[xx]}$ are non-negative for $h\geq 1$ and the activation functions $\tilde{\sigma}_h$ are convex and non-decreasing for $h\geq 0$.
\end{prop}

\medskip

\begin{prop}[\textbf{pICNN -- Type 2}] The pICNN with input $\bx=:\bx_0,\,\by=:\by_0$, output $\cP\left(\bx,\,\by\right):=\bx_{H+1}\in\bbR$ and $H$ hidden layers
    \begin{equation}
        \begin{aligned}\label{simple_passthrough}
            \by_{h+1} &= \sigma_h         \left( \bW_h^{[yy]}\,\by_h + \bb_h^{[y]}\right) &&\in\mathbb{R}^{\mathrlap{n_h}\phantom{m_h}},\quad h=0,\dotsc,H\,,
            \\
            \bx_{h+1} &= \tilde{\sigma}_h \left( \bW_h^{[xx]}\,\bx_h + \bW_h^{[xx_0]}\,\bx_0 + \bW_h^{[xy]}\,\by_h + \bb_h^{[x]} \right)  &&\in\mathbb{R}^{m_h},\quad h=0,\dotsc,H
        \end{aligned}
    \end{equation}
    is convex in $\bx$ given that the weights $\bW_h^{[xx]},\, h \geq 1$ are non-negative and the activation functions $\tilde{\sigma}_h$ are convex and non-decreasing for $h\geq 0$. 
\end{prop}

\medskip

\begin{prop}[\textbf{pICNN -- Type 3}] The pICNN with input $\bx=:\bx_0,\,\by=:\by_0$, output $\cP\left(\bx,\,\by\right):=\bx_{H+1}\in\bbR$ and $H$ hidden layers
    \begin{equation}
        \begin{aligned}\label{eq:type_3}
            \by_{h+1} &= \sigma_h \left( \bW_h^{[yy]}\,\by_h + \bb_h^{[y]} \right) &&\in\mathbb{R}^{\mathrlap{n_h}\phantom{m_h}},\quad h=0,\dotsc,H\,,
            \\
            \bx_{h+1} &= \tilde{\sigma}_h 
            \left( \mathrlap{\bW_h^{[xx]}}\phantom{\bW_h^{[xx_0]}}  \left( \bx_h \hadamard \mathrelu{\tilde{\bW}_h^{\mathrlap{[xy]}\phantom{[x_0y]}}\,\by_h + \mathrlap{\tilde{\bb}_h^{[x]}}\phantom{\tilde{\bb}_h^{[x_0]}}} \right) \;+ \right. \\
            & \phantom{{}=\tilde{\sigma}_h \left(\vphantom{\bW_h^{[xx_0]}}\kern-\nulldelimiterspace\right.}
            \bW_h^{[xx_0]} \left( \mathrlap{\bx_0}\phantom{\bx_h} \hadamard \mathrelu{\tilde{\bW}_h^{[x_0y]}\,\by_h + \tilde{\bb}_h^{[x_0]}} \right) \;+ \\
            & \phantom{{}=\tilde{\sigma}_h \left(\vphantom{\bW_h^{[xx_0]}}\kern-\nulldelimiterspace\right.}
            \bW_h^{[xy]}\by_h + \bb_h^{[x]} \kern-\nulldelimiterspace\left.\vphantom{\bW_h^{[xx_0]}}\right)  &&\in\mathbb{R}^{m_h},\quad h=0,\dotsc,H
        \end{aligned}
    \end{equation}
    is convex in $\bx$ given that the weights $\bW_h^{[xx]},\, h \geq 1$ are non-negative and the activation functions $\tilde{\sigma}_h$ are convex and non-decreasing for $h\geq 0$.
\end{prop}

\medskip

\begin{prop}[\textbf{pICNN -- Type 1M}]\label{prop:monoton} The pICNN with input $\bx=:\bx_0,\,\by=:\by_0$, output $\cP\left(\bx,\,\by\right):=\bx_{H_x+1}\in\bbR$ and $H_x,H_y$ hidden layers
\begin{equation}
    \begin{aligned}
        \by_{h+1} &= \sigma_h        \left(  \bW_h^{[yy]}\,\by_h + \bb_h^{[y]}\right)  &&\in\mathbb{R}^{\mathrlap{n_h}\phantom{m_h}},\quad h=0,\dotsc,H_y\,,
        \\
        \bx_{\mathrlap{1}\phantom{h+1}} &= \tilde{\sigma}_{\mathrlap{1}\phantom{h}}  \left(  \bW_0^{[xx]}\,\bx_0 + \bb_0^{[x]} + \bW^{[xy]}\,\by_{H_y+1} \right)  &&\in\mathbb{R}^{\mathrlap{m_1}\phantom{m_h}}, 
        \\
        \bx_{h+1} &= \tilde{\sigma}_h \left(  \bW_h^{[xx]}\,\bx_h + \bb_h^{[x]} \right)  &&\in\mathbb{R}^{m_h},\quad h=1,\dotsc,H_x
    \end{aligned}
\end{equation}
is convex in $\bx$ and monotonically increasing in $\by$ given that the weights $\bW_h^{[xx]},\,h\geq 1,$ and $\bW_h^{[yy]},\,h\geq 0,$ are non-negative, the activation functions $\tilde{\sigma}_h,\,h\geq 0,$ are convex and non-decreasing, and the activation functions ${\sigma}_h,\,h\geq 0,$ are non-decreasing. If at least one activation function ${\sigma}_h$ is not convex, the pICNN is not convex in $\by$.
\end{prop}

\medskip

\medskip

To construct pICNNs which are convex and \emph{non-decreasing} in $\bx$, also the weights acting directly on $\bx$ must be non-negative. This means that for all types, $\bW_h^{[xx]}$ has to be non-negative for all $h$. 
Both Type 2 and Type 3 use so called passthrough layers, which pass the argument $\bx$ into every hidden layer. In conventional (p)ICNNs, passthrough layers have a significant benefit. Here, the NN must not necessarily be non-decreasing in the input, as naturally, convex functions can also be decreasing, cf. \cref{sec:1d}. Thus, the weights acting directly on the input may take positive or negative values. Using passthrough layers exploits this benefit in every layer of the NN. However, as in the application to polyconvexity, also the weights of the passthrough layer $\bW_h^{[xx_0]}$ must be non-negative, their benefit is limited.

Furthermore, as only the gradient of the potential is considered in this work, cf. Eq.~\eqref{eq:stress}, all contributions to the output layer which are independent of the invariants are omitted, such as the bias in the output layer and the last two parameter layers of Type 2.

Throughout this work, the convex and non-decreasing {Softplus} activation function, cf. \cref{fig:1d}, is applied in all hidden layers for both $\sigma_h$ and $\tilde{\sigma}_h$, except for Type 1M, where the monotonically increasing but non-convex {Sigmoid} activation function is applied in the first layer of the parameter input. In the output layer, a linear activation function is applied. By this, the potential is infinitely continuously differentiable in $\bx$. Type 1 and Type 2 are also infinitely continuously differentiable in $\by$. However, due to the application of the ReLu function in Type 3, this architecture is not continuously differentiable in $\by$. This could be circumvented by applying any positive and continuously differentiable function instead of ReLu, e.g., the Softplus function. Note again that the adapted architectures are visualized in \cref{fig:pICNN architectures} for one specific choice of nodes and layers. 

\section{Numerical examples}\label{sec:example}

\subsection{Scalar-valued parametrisation}

\subsubsection{Data generation}\label{sec:data_example_one}

\begin{figure}[t!]
    \centering
    \begin{subfigure}[b]{0.475\textwidth}
        \centering
        \resizebox{\textwidth}{!}{
        \begin{tikzpicture}
\begin{axis}[
        xlabel=$t$,
        ylabel=$\mu(t)$,
        xmin=0, xmax=1, % x scale
        ymin=0.3, ymax=2.7, % y scale
        domain=0:1,   % added, key improvements
        legend style={at={(0.5,-0.175)},anchor=north,font=\large},
        legend columns = 4
        ]
    \addplot [2darkblue,mark = triangle*, mark repeat = 2, mark size = 2.5pt,ultra thick]    {0.5 + 2*x};
 %   \addlegendentry{Case A}
    \addplot [2red, ultra thick, mark=*,mark repeat = 2, mark size = 2.5pt]    {8*x^2-8*x+2.5};
 %   \addlegendentry{Case B}
    \addplot [2orange,ultra thick, mark=square*,mark repeat = 2, mark size = 2.5pt]    {-8*x^2+8*x+0.5};
 %   \addlegendentry{Case C}

   %     \node[left,text=2green] at (-0.01,2.5) {$\large\boldsymbol{\text{B}}$};   
%        \node[left,text=2darkblue] at (-0.01,1) {$\large\boldsymbol{\text{C}}$};   
%        \node[left,text=2red] at (-0.01,0.5) {$\large\boldsymbol{\text{A}}$};   

    \addlegendentryexpanded{A}
    \addlegendentryexpanded{B}
    \addlegendentryexpanded{C}

\end{axis}
\end{tikzpicture}
        }
        \caption{}
    \end{subfigure}
    \begin{subfigure}[b]{0.475\textwidth}   
        \centering 
        \resizebox{\textwidth}{!}{
        \begin{tikzpicture} 
\begin{axis} [
        xlabel=$t$,
        ylabel=$\lambda(t)$,
        xmin=0, xmax=1, % x scale
        ymin=98.2 , ymax=99.8,
        domain=0:1,
        legend style={at={(0.5,-0.175)},anchor=north,font=\large},
                        legend columns = 4
        ]
    \addplot [2darkblue,mark=triangle*, mark repeat = 2, mark size = 2.5pt,ultra thick]{100-2/3*(0.5 + 2*x)};
    \addplot [2red, ultra thick, mark=*,mark repeat = 2, mark size = 2.5pt]    {100-2/3*(8*x^2 -8*x +2.5)};
    \addplot [2orange,ultra thick, mark=square*,mark repeat = 2, mark size = 2.5pt] {100-2/3*(-8*x^2 +8*x +0.5)};

     %   \node[left,text=2green] at (-0.01,98.33) {$\large\boldsymbol{\text{B}}$};   
     %   \node[left,text=2darkblue] at (-0.01,99.4) {$\large\boldsymbol{\text{C}}$};   
     %   \node[left,text=2red] at (-0.01,99.67) {$\large\boldsymbol{\text{A}}$};   

    \addlegendentryexpanded{A}
    \addlegendentryexpanded{B}
    \addlegendentryexpanded{C}

\end{axis} 
\end{tikzpicture}
        }    
        \caption{}
    \end{subfigure}
    \caption{Three different parametrisations (A-C) of the Lam\'e constants in the Neo-Hookean model.}
    \label{fig:parameterisation}
\end{figure}
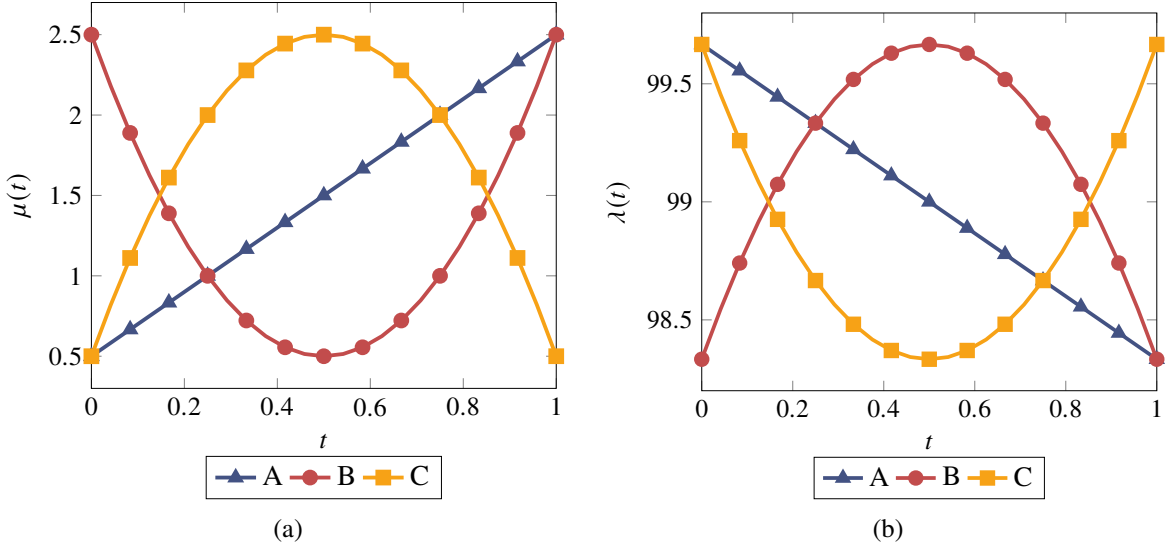

As a first proof of concept, the models proposed in \cref{sec:model} are calibrated to data generated with the parametrised Neo-Hookean potential introduced in Eq.~\eqref{eq:NH_param}. For this, three different parametrisations 
\begin{equation}
    {\mu}(t) = \begin{cases}
			0.5 + 2t      , & \text{Case A}\\
            8t^{2} -8t +2.5       , & \text{Case B}\\
            -8t^{2} +8t +0.5      , & \text{Case C}
		 \end{cases}, \\
	\qquad {\lambda}(t) = {\kappa} - \frac{2}{3}{\mu}(t), \qquad t \in [0,1] \,,
\end{equation}
of the Lam\'e parameters $\mu,\lambda$ with a constant bulk modulus ${\kappa} = 100$ are applied.
The different parametrisations are chosen such that the hyperelastic potential has both convex and concave dependencies on the parameter $t$, cf.~\cref{fig:parameterisation}. Thus, the pICNN Types 1--3 are examined, meaning the architectures which represent arbitrary functional relationships in the parameter.
Overall, discrete values for both the deformation gradient $\bF$ and the scalar parameter $t$ have to be sampled for the data generation, resulting in datasets of the form
\begin{equation}
    \mathcal{D}=\left\{\big(\: ^1\bF,\,^1t;\,^1\bP\big),\dotsc \big(\: ^n\bF,\,^nt;\,^n\bP\big)\right\}\,,
\end{equation}
where in each tuple, the prescribed deformation gradient $^i\bF$ and the parameter $^it$ have a corresponding first Piola-Kirchhoff stress $^i\bP$. As the data is generated with an analytical potential, also the values of the potential $\psi(^i\bF,^it)$ are available and could be included in the dataset. However, as real-world experiments only provide stress values, this would be a less general approach. Also, even when data of the potential is available, including it in the calibration process does barely improve the model quality \cite{klein2022a}. 
Thus, the potential is calibrated only through its gradients, which is referred to as Sobolev training \cite{vlassis}.

Following \cite{fernandez}, the sampling of the stress-strain states is motivated by physical experiments which could also be applied in experimental investigations. In particular, a uniaxial tension stress state, a biaxial tension stress state, and a shear deformation state are applied, where each load case consists of 101 datapoints and the data is generated by numerically solving the underlying equation systems for each load case. 
The uniaxial tension is applied in $x$-direction with $F_{11}\in[0.5,\,1.5]$, the equibiaxial tension is applied in $x-y$-direction with $F_{11}=F_{22}\in[0.5,\,1.5]$, and simple shear is applied with $F_{12}\in[-0.5,\,0.5]$.
In addition, a mixed shear-tension test is applied, which represents a fairly general deformation mode ("test 3" in \cite{fernandez}). The parameter $t\in [0,\,1]$ is sampled with 201 equidistant points.

\subsubsection{Model preparation and calibration}

In this example, the pICNN architectures with an arbitrary functional relationship in the parameter $t$ are applied, i.e., Type 1--3.
The hyperparameters, i.e., number of nodes and layers, of the different pICNN architectures described in \cref{sec:architectures} are chosen such that they are in the same order of magnitude for all models. The number of nodes and layers are visualized in \cref{fig:pICNN architectures}. The total number of trainable parameters for the models using the Type 1--3 pICNNs are 272, 516, and 580, respectively.
For the model calibration, the loss function given as the mean squared error
\begin{equation}
    \text{MSE}=\frac{1}{9lmn}\sum_{i=1}^l\sum_{j=1}^m\frac{1}{w_{ij}}\sum_{k=1}^n\norm{^{ijk}\bP-\bP\left(^{ik}\bF;\,^j t\right)}^2
\end{equation}
is minimized. Here, the outer loop over $i$ corresponds to the $l$ load paths in the calibration dataset. Each load path is combined with $m$ different, fixed $t$ values, where the sum over $j$ corresponds to the values of $t$. Finally, for one fixed combination of load path and parameter $t$, the weight is calculated according to the norm
\begin{equation}
    w_{ij}=\frac{1}{n}\sum_{k=1}^n\norm{^{ijk}\bP}\,,
\end{equation}
and the innermost sum over $k$ corresponds to the $n$ different deformation gradients. 
For the evaluation of the loss after the model calibration, all weights are set to one, i.e., $w_{ij}=1$.
The models are implemented in TensorFlow 2.10.0, using Python 3.10.9. For the optimization, the Adam optimizer is used with a learning rate of 0.002 and $7,000$ epochs.  The full batch of training data is used with TensorFlow’s default batch size.

\paragraph{Study I}
For the calibration dataset, the uniaxial, biaxial and shear loads are combined with $t\in\{0,\,0.2,\,\allowbreak 0.4,\,0.6,\,0.8,\,1\}$. Thus, the calibration dataset consists of $1,818$ tuples. For the test dataset, the mixed shear-tension test is combined with all remaining 195 values for $t$ not included in the calibration, thus consisting of $19,695$ tuples. 
For this study, the PANN model as described in \cref{sec:model} is applied, with three different versions using the pICNN architectures Type 1--3 as described in \ref{sec:architectures}. Each model is calibrated five times to each parametrisation case, where the model with the worst test loss is exluded.

\paragraph{Study II}
For the calibration dataset, the uniaxial, biaxial and shear tension loads are combined with $t\in\{0,\,0.1,\,0.9,\,1\}$, yielding a calibration dataset with $1,212$ tuples. 
For the test dataset, the mixed shear-tension test is combined with the 197 remaining values for $t$ not included in the calibration, thus consisting of $19,897$ tuples.
For this study, the model as described in \cref{sec:model} is adapted in such a way that it does not include the normalisation term \cref{eq:norm}. Here, the pICNN architecture Type 1 is applied. One model instance with the normalisation condition, and one without is calibrated. The model is calibrated one time to the parametrisation case A.

\subsubsection{Results}

\begin{figure}[t!]
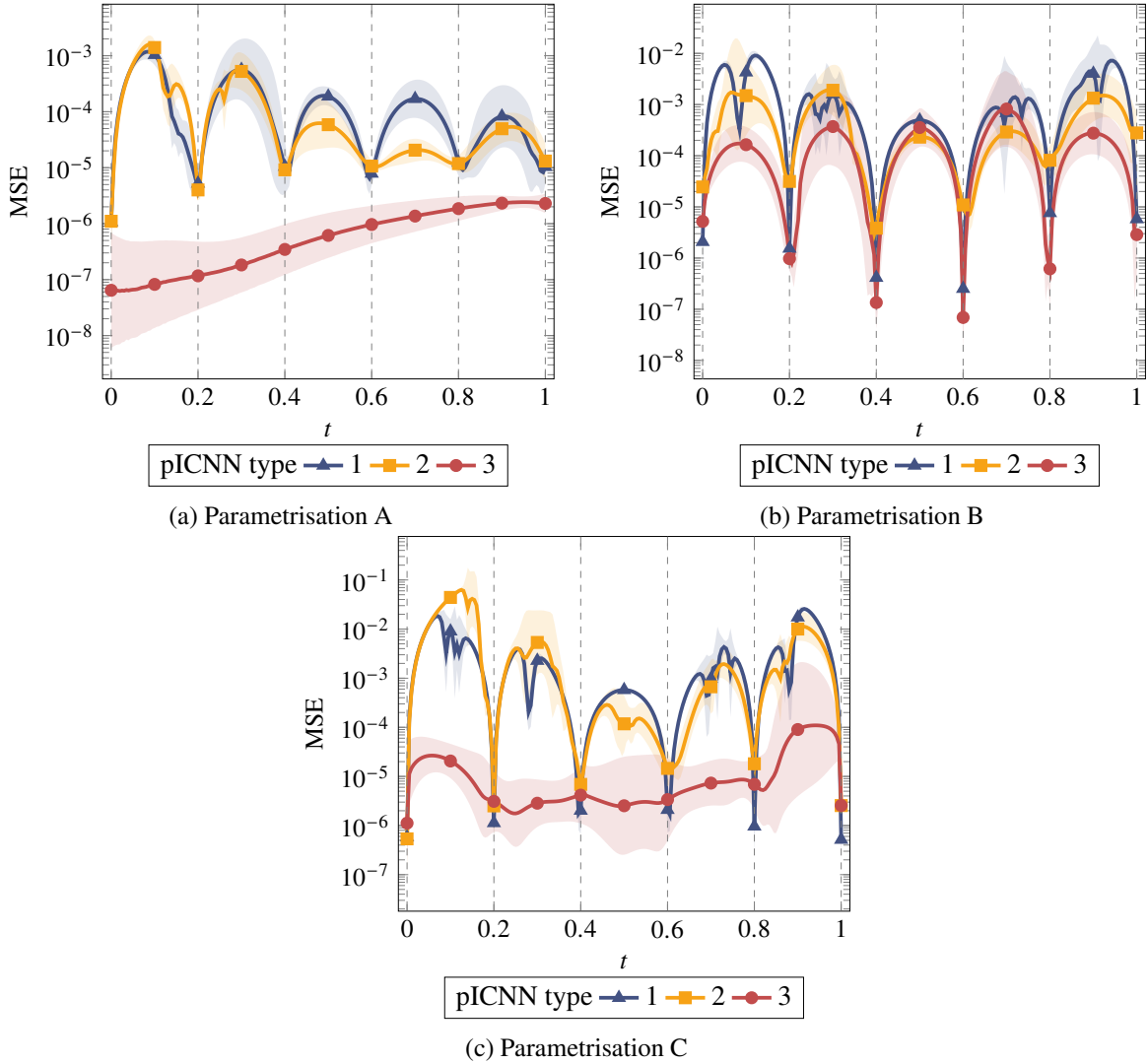

    % Depicted:
    % The test3 prediction MSE over the parameter-domain for the four best performing models (the worst performing instance was removed (worst performing = worst overall MSE over all loadcases and parameters))
    \centering
    \begin{subfigure}[b]{0.475\textwidth}
        \centering
        \resizebox{\textwidth}{!}{
            \MSEtplot{data/A_equi_test3_mean_t_losses.txt}
        }
        \caption{Parametrisation A}
        \label{fig:average MSE Case A}
    \end{subfigure}
    \begin{subfigure}[b]{0.475\textwidth}   
        \centering 
        \resizebox{\textwidth}{!}{
            \MSEtplot{data/B_equi_test3_mean_t_losses.txt}
        }    
        \caption{Parametrisation B}
        \label{fig:average MSE Case B}
    \end{subfigure}
    \begin{subfigure}[b]{0.475\textwidth}   
        \centering 
        \resizebox{\textwidth}{!}{
            \MSEtplot{data/C_equi_test3_mean_t_losses.txt}
        }    
        \caption{Parametrisation C}
        \label{fig:average MSE Case C}
    \end{subfigure}
    \caption{Continuous line denotes the average of $\log_{10}$ MSE, while shaded areas denote the standard deviation of $\log_{10}$ MSE.}
    \label{fig:average MSE}
\end{figure}

\begin{figure}[t!]
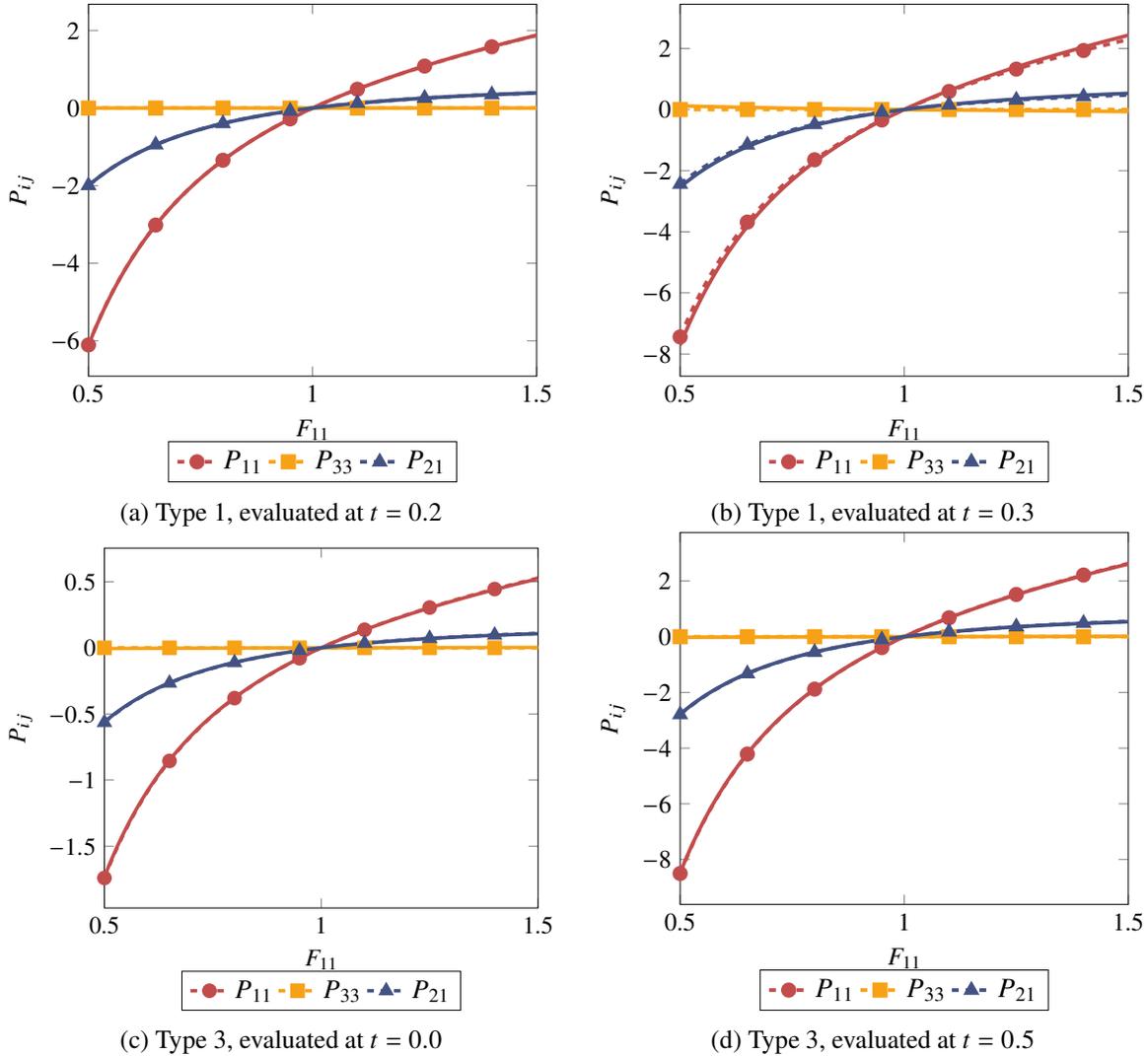
\label{fig:stress1}
    % Depicted:
    % The test3 prediction for Type 1 and Type 3 (trained on the 6 equidistant t values) for different t values, in all cases the first (equivalent to randomly chosen) instance was used (here no outlier removal was performed!). The log10(MSE) for the specific predictions are currently printed in the caption. Parametrisation used: case C
    \centering

    \begin{subfigure}[b]{0.475\textwidth}   
        \centering 
        \resizebox{\textwidth}{!}{
            \predictionplot{data/C_equi_Type_A_Inst0_t0200.txt}
        }    
        \caption{Type 1, evaluated at $t=0.2$}
        \label{fig:prediction Type 1 t02}
    \end{subfigure}
    \begin{subfigure}[b]{0.475\textwidth}   
        \centering 
        \resizebox{\textwidth}{!}{
            \predictionplot{data/C_equi_Type_A_Inst0_t0300.txt}
        }    
        \caption{Type 1, evaluated at $t=0.3$ }
        \label{fig:prediction Type 1 t03}
    \end{subfigure}
        \begin{subfigure}[b]{0.475\textwidth}   
        \centering 
        \resizebox{\textwidth}{!}{
            \predictionplot{data/C_equi_Type_D_Inst0_t0000.txt}
        }    
        \caption{Type 3, evaluated at $t=0.0$}
        \label{fig:prediction Type 3 t00}
    \end{subfigure}
    \begin{subfigure}[b]{0.475\textwidth}   
        \centering 
        \resizebox{\textwidth}{!}{
            \predictionplot{data/C_equi_Type_D_Inst0_t0500.txt}
        }    
        \caption{Type 3, evaluated at $t=0.5$}
        \label{fig:prediction Type 3 t05}
    \end{subfigure}
    \caption{Results for parametrisation case C, evaluated for the mixed shear-tension test case. Dashed lines and points denote the data, while continuous lines denote the model prediction.
}
\label{fig:study1_P}
\end{figure}

\begin{table}[t]
     \centering
\tabcolsep=0pt%
{
    \centering
    \begin{tabular*}{0.8\textwidth}{@{\extracolsep{\fill}}lcccccc@{}}\toprule%
    & \multicolumn{3}{@{}c@{}}{{Calibration}}& \multicolumn{3}{@{}c@{}}{{Test}} \\\cmidrule{2-4}\cmidrule{5-7}%
    {} & {Case A} & {Case B} & {Case C} & {Case A} & {Case B} & {Case C} \\\midrule
    Type 1 & -4.54 & -4.44 & -5.24 & -3.60 & -2.69 & -2.43 \\
    Type 2 & -4.55 & -3.26 & -4.31 & -3.70 & -2.96 & -2.12 \\
    Type 3 & -5.42 & -4.74 & -4.13 & -5.97 & -3.45 & -3.61 \\
    \bottomrule
    \end{tabular*}%
}
\caption{
%\TBL{\caption{
Average $\log_{10}$ MSE for the scalar-valued parametrisation. Four best calibrated model instances for study I for all parametrisation cases and pICNN types.
 \label{tab:interpolation extrapolation MSE outlier removed}}
\end{table}

\begin{figure}[t!]
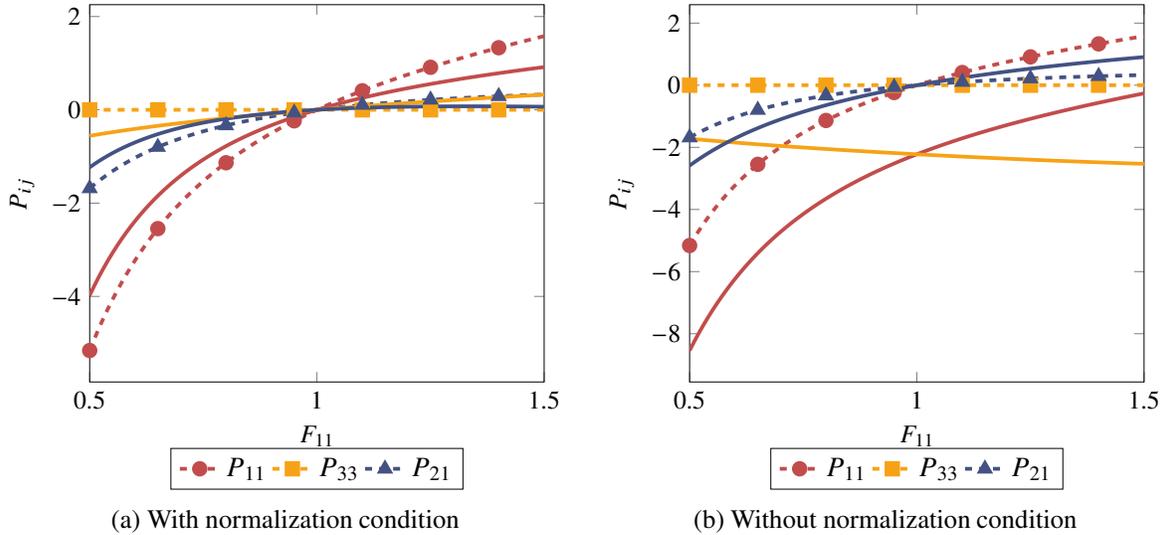


    % Depicted:
    % Stress prediction of first instance of Type 1 trained on t in [0.0, 0.1, 0.9, 1.0] and evaluated at t=0.5. Left with and right without normalization term. Parametrization used: Case A

    \centering
    \begin{subfigure}[b]{0.475\textwidth}
        \centering
        \resizebox{\textwidth}{!}{
            \predictionplot{data/A_edges_Type_A_Inst0_t0500.txt}
        }
        \caption{With normalization condition}
        \label{fig:with normalization}
    \end{subfigure}
    \begin{subfigure}[b]{0.475\textwidth}   
        \centering 
        \resizebox{\textwidth}{!}{
            \predictionplot{data/A_edges_Type_A_nonnormalized_t0500.txt}
        }    
        \caption{Without normalization condition}
        \label{fig:without normalization}
    \end{subfigure}
    \caption{Results for parametrisation case C, evaluated for the mixed shear-tension case. In this case, the model was only calibrated on the edges of the parameter domain of $t$, and evaluated in the middle. Dashed lines and points denote the data, while continous lines denote the model prediction.}
    \label{fig:stress_study_2}
\end{figure}

\paragraph{Study I}

In \cref{tab:interpolation extrapolation MSE outlier removed}, the MSE values of the calibrated models are presented for all pICNN architectures and all parametrisation cases. In general, all pICNN architectures are able to interpolate the data for all parametrisation cases, and also show excellent performance on the test dataset for all parametrisations. While for the calibration, the pICNN Type 3 performs slightly better, it performs way better on the test dataset than the remaining architectures. 
However, due to the simplicity of the examined data, no premature conclusions about the general performance of the different architectures should be drawn. Even the architecture with the lowest complexity might be sufficiently flexible in practical applications.
The MSEs evaluated for the mixed shear-tension load and all values for $t$ is visualized in \cref{fig:average MSE}. Not surprisingly, the models perform better for values of $t$ which were included in the model calibration. In particular for parametrisation A and C, the pICNN Type 3 performs way better than the other architectures. Note that, when leaving the training values of $t$, the MSE increases quite quickly for Type 1 and 2, which could be a sign of overfitting in the parameter $t$. For parametrisation B, the pICNN architecture 3 shows a similar behavior compared to the other pICNN architectures. While pICNN Type 1 and 2 have quite similar prediction qualities between different calibration instances, pICNN Type 3 has a higher discrepancy between different calibration instances, as indicated by the shaded areas. 

For the following investigation, a random instance of the four trained models was chosen.
In \cref{fig:study1_P}, some stress predictions of the models are visualized for the parametrisation case C. On the top row, a model using a Type 1 pICNN is evaluated, while on the bottom row, a model using a Type 3 pICNN is evaluated. On the left hand side, the models are evaluated for values of $t$ used in the calibration, while on the right hand side, the models are evaluated for values of $t$ not used in the calibration. In all cases, the model has to extrapolate in the load case. The interpolation is excellent for both evaluated models. For Type 1, the evaluation for $t=0.3$ shows some visible deviations from the ground truth. This case has a $\log_{10}$ MSE of -2.65. Thus, with the MSEs of \cref{fig:average MSE} in mind, this is a representative case for the less good model predictions. And still, the prediction quality might be good enough for most practical applications. The pICNN Type 3 is able to also perfectly make predictions at $t=0.5$, although the magnitude of the stress components differs by a factor of $\approx$ 4 for different values of $t$.

\paragraph{Study II}

In \cref{fig:stress_study_2}, a comparison is made between a model which fulfills the normalisation condition of Eq.~\eqref{eq:norm} by construction and one which only learns to approximate the condition through the calibration dataset. This case includes values of $t$ only on the edges of its parameter domain and is here evaluated for the value in the middle. While for the model which fulfills the normalisation condition by construction, there are some significant deviations from the ground truth, it still fulfills the normalisation condition in an exact way. The model which does not fulfill the normalisation condition by construction has to learn it by the data which fulfill this property. While this indeed works out for the values of $t$ included in the calibration dataset, for the case visualized here it is violated quite obviously. This example demonstrates the benefit of fulfilling mechanical conditions by construction, in particular for the generalization of the model.

\subsection{Vector-valued parametrisation with monotonicity condition}

\subsubsection{Data generation}
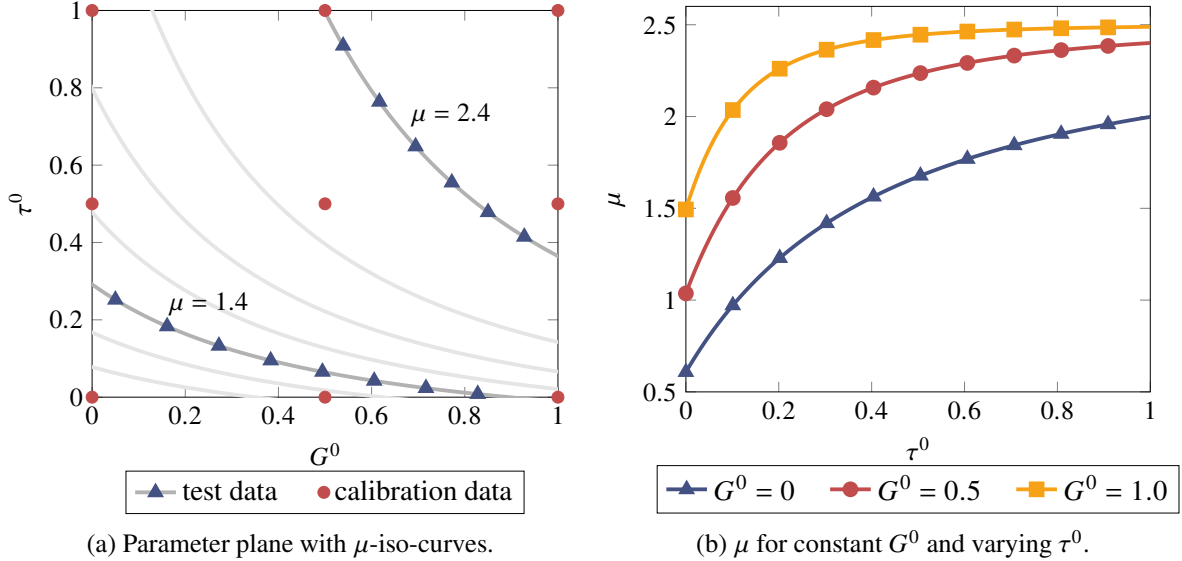
\begin{figure}[t!]
    \centering
    \begin{subfigure}[b]{0.475\textwidth}
        \centering
        \resizebox{\textwidth}{!}{
        \begin{tikzpicture}
\begin{axis}[
        xlabel=$G^0$,
        ylabel=$\tau^0$,
        xmin=0, xmax=1, % x scale
        ymin=0, ymax=1, % y scale
        domain=0:1,   % added, key improvements
        legend style={at={(0.5,-0.19)},anchor=north,font=\large},
        legend columns = 4
        ]

            %      \addlegendimage{empty legend}
   % \addlegendentryexpanded{$\mu$}

 %   \addplot [only marks, mark=triangle*,2darkblue,mark size=3.5pt]
 % table[row sep=crcr]{%
%0.1 0.218 \\
%0.3 0.12233 \\
%0.5 0.06422 \\
%0.6 0.79209 \\
%0.75 0.57975 \\
%0.9 0.4361 \\
%};

  \addplot[2grey!80!black,mark options ={2darkblue}, mark = triangle*, mark repeat = 10, mark size = 2.5pt,ultra thick, mark phase = 5] table[x index=0,y index=1] {data/print_data/14.txt};

    \addlegendentryexpanded{test data $\quad$}

    \addplot [only marks, mark=*,2red,mark size=2.5pt]
  table[row sep=crcr]{%
0 0 \\
0 0.5 \\
0 1 \\
0.5 0 \\
0.5 0.5 \\
0.5 1 \\
1 0 \\
1 0.5 \\
1 1 \\
};

    \addlegendentryexpanded{calibration data}

  \addplot[2grey!80!black,mark options ={2darkblue}, ultra thick, mark=triangle*,mark repeat =7, mark size = 2.5pt, mark phase = 4] table[x index=0,y index=1] {data/print_data/24.txt};

%\addplot[2grey!80!black,mark = triangle*, mark repeat = 10, mark size = 2.5pt,ultra thick] table[x index=0,y index=1, mark=none] {data/print_data/14.txt};
%\addplot[2grey!80!black, ultra thick, mark=*,mark repeat = 10, mark size = 2.5pt] table[x index=0,y index=1,mark=none] {data/print_data/24.txt};

\addplot[2grey!80, ultra thick] table[x index=0,y index=1,mark=none] {data/print_data/09.txt};
\addplot[2grey!80, ultra thick] table[x index=0,y index=1,mark=none] {data/print_data/115.txt};
\addplot[2grey!80, ultra thick] table[x index=0,y index=1,mark=none] {data/print_data/165.txt};
\addplot[2grey!80, ultra thick] table[x index=0,y index=1,mark=none] {data/print_data/19.txt};
\addplot[2grey!80, ultra thick] table[x index=0,y index=1,mark=none] {data/print_data/215.txt};

      %  \addplot [2darkblue,mark = triangle*, mark repeat = 2, mark size = 2.5pt,ultra thick]    {0.5 + 2*x};
 %   \addlegendentry{Case A}
   % \addplot [2red, ultra thick, mark=*,mark repeat = 2, mark size = 2.5pt]    {8*x^2-8*x+2.5};
 %   \addlegendentry{Case B}
   %\addplot [2orange,ultra thick, mark=square*,mark repeat = 2, mark size = 2.5pt]    {-8*x^2+8*x+0.5};
 %   \addlegendentry{Case C}

  %      \addplot [2darkblue,mark = triangle*, mark repeat = 2, mark size = 2.5pt,ultra thick]    {0.5 + 2*x};

   %     \node[left,text=2green] at (-0.01,2.5) {$\large\boldsymbol{\text{B}}$};   
%        \node[left,text=2darkblue] at (-0.01,1) {$\large\boldsymbol{\text{C}}$};   
%        \node[left,text=2red] at (-0.01,0.5) {$\large\boldsymbol{\text{A}}$};   

\node at (0.25,0.24) {$\mu=1.4$};
\node at (0.77,0.73) {$\mu=2.4$};

  %  \addlegendentryexpanded{$1.0$}
  %  \addlegendentryexpanded{$1.6$}
  %  \addlegendentryexpanded{$2.2$}

\end{axis}
\end{tikzpicture}
        }
        \caption{Parameter plane with $\mu$-iso-curves.}
    \end{subfigure}
    \begin{subfigure}[b]{0.48\textwidth}   
        \centering 
        \resizebox{\textwidth}{!}{
        \begin{tikzpicture}
\begin{axis}[
        xlabel=$\tau^0$,
        ylabel=$\mu$,
        xmin=0, xmax=1, % x scale
        ymin=0.5, ymax=2.6, % y scale
        domain=0:1,   
     %   legend pos = south east% added, key improvements
        legend style={at={(0.5,-0.19)},anchor=north,font=\large},
        legend columns = 4
        ]

                      %\addlegendimage{empty legend}
    %\addlegendentryexpanded{$G^0$}

% \addplot[2lightblue, ultra thick] table[x index=0,y index=1,mark=none] {data/print_data/G_0.txt};
% \addplot[2darkblue, ultra thick] table[x index=0,y index=1,mark=none] {data/print_data/G_05.txt};
% \addplot[2red, ultra thick] table[x index=0,y index=1,mark=none] {data/print_data/G_1.txt};

    \addplot[2darkblue,mark = triangle*, mark repeat = 10, mark size = 2.5pt,ultra thick] table[x index=0,y index=1] {data/print_data/G_0.txt};
\addplot[2red, ultra thick, mark=*,mark repeat = 10, mark size = 2.5pt] table[x index=0,y index=1] {data/print_data/G_05.txt};
\addplot[2orange,ultra thick, mark=square*,mark repeat = 10, mark size = 2.5pt] table[x index=0,y index=1] {data/print_data/G_1.txt};

  %      \addplot [2darkblue,mark = triangle*, mark repeat = 2, mark size = 2.5pt,ultra thick]    {0.5 + 2*x};

   %     \node[left,text=2green] at (-0.01,2.5) {$\large\boldsymbol{\text{B}}$};   
%        \node[left,text=2darkblue] at (-0.01,1) {$\large\boldsymbol{\text{C}}$};   
%        \node[left,text=2red] at (-0.01,0.5) {$\large\boldsymbol{\text{A}}$};   

    \addlegendentryexpanded{$G^0=0\quad$}
    \addlegendentryexpanded{$G^0=0.5\quad$}
    \addlegendentryexpanded{$G^0=1.0$}

\end{axis}
\end{tikzpicture}
        }    
        \caption{$\mu$ for constant $G^0$ and varying $\tau^0$.}
    \end{subfigure}
    \caption{Dependency of the shear modulus $\mu$ on the  vector-valued, 3D printing-inspired  parametrisation in terms of $(G^0,\tau^0)$.}
    \label{fig:parameterisation_vector}
\end{figure}

In the next example, the Neo-Hookean potential
\begin{equation}\label{eq:NH_param_2}
     \psi^{\text{nh}}(I_1,I_3; \,\bt) = \frac{\mu(\bt)}{2}\left(I_1 - 3 - 2\ln \sqrt{I_3}\right) + \frac{\lambda}{2}\left(\sqrt{I_3}-1\right)^2\,
 \end{equation}
is considered, where $\mu(\bt)$ is parametrised in $\bt=\left(G^0,\,\tau^0\right)\in[0,1]^2$ and $\lambda=100=\text{const}$. The parametrisation
\begin{equation}\label{eq:param_2}
\begin{aligned}
    &{\mu}(\bt) =2.5 \,\operatorname{tanh}H^v\,,\qquad &&H^v=1.7\,\hat{G}^2 \operatorname{ln}  \hat{\tau}\,,
	\\
	   & \hat{G}=0.6+0.4 G^0\,,\qquad && \hat{\tau}=1.5+4.5\tau^0\,,
	   %\\
	   % &G^0,\tau^0\in[0,1]\,,
	\end{aligned}
\end{equation}
is inspired by a 3D printing process, where a liquid photopolymer resin is hardened by exposing it to ultraviolet light for a given time, cf.~\cite{valizadeh}. 
The properties of the final solid material depend on both the light intensity, which is determined through the greyscale value $\hat G$, and the time $\hat \tau$ the light is applied on the resin.
Here, these two parameters are parameterised in a physically sensible range through $G^0$, which is associated with the light intensity, and $\tau^0$, which is associated with the exposure time for which the light is applied on the resin.
The parametrisation consists of three ideas. First of all, the shear modulus $\mu$ is influenced by both $G^0$ and $\tau^0$. In particular, the same shear modulus can be achieved by different combinations of $(G^0,\,\tau^0)$. This is reflected by the intermediate quantity $H^v$, cf.~Eq.~\eqref{eq:param_2}. Secondly, the shear modulus is bounded from above, which is reflected by the $\operatorname{tanh}$ function which receives $H^v$ as an input. Lastly, the shear modulus is a monotonically increasing function in $(G^0,\,\tau^0)$, which reflects the physical observation that the shear modulus increases when increasing the light intensity or the light exposure time.
In \cref{fig:parameterisation_vector}, these characteristics are visualized. 

The deformation gradients are sampled as described in \cref{sec:data_example_one}. For the calibration dataset, 9 parameter combinations $\bt=(G^0,\,\tau^0)$ are sampled, cf.~\cref{fig:parameterisation_vector}. For the test dataset, two $\mu$-iso-curves are considered for $\mu\in\{1.4,\,2.4\}$, cf.~\cref{fig:parameterisation_vector}. For each iso-curve, 100 $(G^0,\,\tau^0)$ tuples are sampled. Overall, this results in a calibration dataset with $2,727$ tuples and a test dataset with $20,200$ tuples.

%
%With the parameters $a=2.5,\,b=1.7$. 

\subsubsection{Model preparation and calibration}

In this example, the pICNN architecture with a monotonically increasing functional relationship in the parameter $\bt$ is applied, i.e., Type 1M. The number of nodes and layers are visualized in \cref{fig:pICNN architectures}, where the total number of trainable parameters is 280.
For the model calibration, all stress values are normalised by the inverse mean Frobenius norm of all tuples in the calibration dataset. Then, the loss function given as the mean squared error
\begin{equation}
    \text{MSE}=\frac{1}{9n}\sum_{i=1}^n\frac{1}{w_i}\norm{^i\bP-\bP\left(^i\bF;\,^i\bt\right)}^2
\end{equation}
is minimized, where $n$ is the number of tuples in the calibration dataset. The sample weight
\begin{equation}
    w_i=\norm{^i\bP} + 1 \geq 1
\end{equation}
is calculated for each single tuple and used to encourage better accuracy of the model when predicting small stress values. For the optimization, the SLSQP optimizer is used.

\subsubsection{Results}
\begin{figure}[t!]
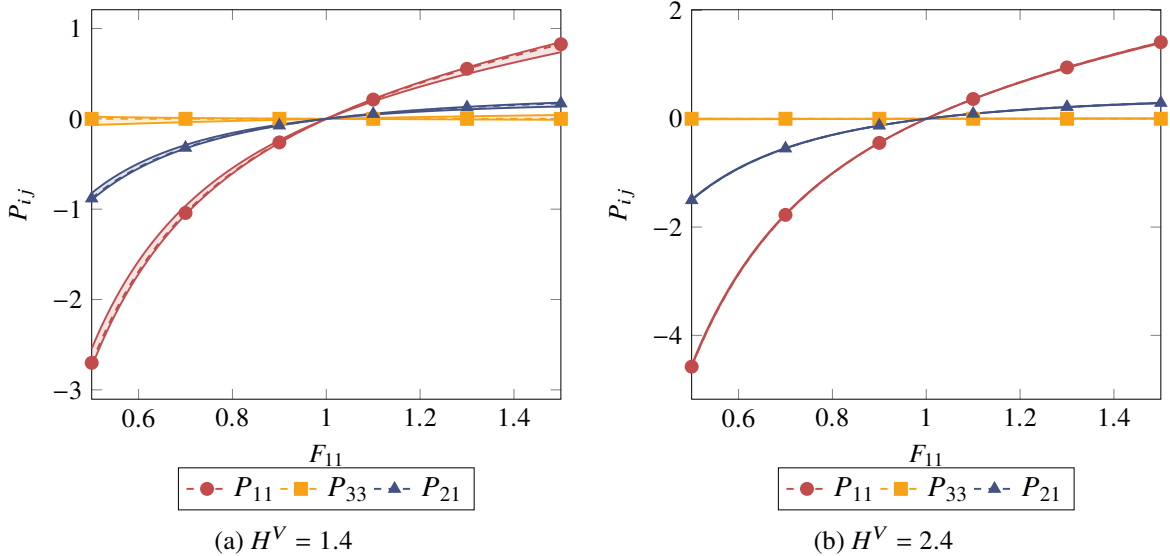

    % Depicted:
    % The test3 prediction for Type 1 and Type 3 (trained on the 6 equidistant t values) for different t values, in all cases the first (equivalent to randomly chosen) instance was used (here no outlier removal was performed!). The log10(MSE) for the specific predictions are currently printed in the caption. Parametrisation used: case C
    \centering

    \begin{subfigure}[b]{0.475\textwidth}   
        \centering 
    \end{subfigure}
        \begin{subfigure}[b]{0.475\textwidth}   
        \centering 
        \resizebox{\textwidth}{!}{
            \StressAreaPlot{data/print_data/vector_param/pIMCNN_area_plot_iso_1_4.txt}
        }    
        \caption{$H^V=1.4$}
    \end{subfigure}
    \begin{subfigure}[b]{0.475\textwidth}   
        \centering 
        \resizebox{\textwidth}{!}{
            \StressAreaPlot{data/print_data/vector_param/pIMCNN_area_plot_iso_2_4.txt}
        }    
        \caption{$H^V=2.4$}
    \end{subfigure}
    \caption{Model prediction for the mixed shear-tension test case. Dashed lines and points denote the data, while lines and shaded areas depict the calibrated model evaluated for different parameter combinations $(G^0,\,\tau^0)$ on $H^v$ iso-curves.}
    \label{fig:stress_print}
\end{figure}

In \cref{tab:MSE_print_data}, the average $\log_{10}$ MSE of the four model instances with the best test MSE, as well as the $\log_{10}$ MSE of the model instance with the best test MSE are presented. The performance on both the calibration and the test dataset is excellent. In \cref{fig:stress_print}, the stress predictions for the test case for $H^v\in\{1.4,\,2.4\}$ are visualized for the best model instance. The calibrated model is evaluated for 100 different $(G^0,\,\tau^0)$ combinations on each $\mu$-iso-curve. For $H^v=2.4$, the model perfectly learns the invariance of $H^v$ in $(G^0,\,\tau^0)$. Thus, the model predictions for different $(G^0,\,\tau^0)$ combinations is practically identical. For $H^v=1.4$, the model predictions slightly differ for different $(G^0,\,\tau^0)$ combinations, which is indicated by the (fairly small) shaded areas in \cref{fig:stress_print}. Overall, the model performs excellent, in particular with regard to the low amount of $(G^0,\,\tau^0)$ samples in the calibration dataset.

\begin{table}[t]
    \centering
\tabcolsep=0pt%
    \begin{tabular*}{0.5\textwidth}{@{\extracolsep{\fill}}lcc@{}}\toprule%
    & {Calibration}& {Test}  \\\midrule
    Average of best four & -4.37 & -3.23  \\
    Best model   & -5.18 & -3.69  \\
    \bottomrule
    \end{tabular*}%
\caption{
Average $\log_{10}$ MSE for the vector-valued parametrisation. Four best calibrated model instances.
 \label{tab:MSE_print_data}}
\end{table}

\section{Conclusion}\label{sec:conc}

In the present work, a NN-based constitutive model for parametrised hyperelasticity is proposed. The model is formulated in such a way that it fulfills all common constitutive conditions of hyperelasticity by construction, without being too restrictive in the parametric dependencies of the model. In particular, by applying partially input-convex neural networks (pICNN), the model fulfills the polyconvexity condition, while still being able to represent arbitrary functional relationships in the additional parameter. In addition, a polyconvex potential is proposed which is monotonic in the additional parameters. 

As a first proof of concept, the model is calibrated to data generated with an analytical potential which depends on one scalar-valued parameter. Different pICNN architectures with different complexities are examined, where all architectures performed excellent. However, due to the simplicity of the examined data, no premature conclusions about the general performance of the different architectures should be drawn. Even the architecture with the lowest complexity might be sufficiently flexible in practical applications.
Furthermore, the proposed model is calibrated to data generated with an analytical potential which depends on multiple parameters. In this case, the dependency of the ground truth potential in the parameters is monotonic, and the NN-based potential which by construction is monotonic in the additional parameters is applied. Again, the model shows excellent performance.

The extension of the proposed framework to multiphysical constitutive models, such as electro-elasticity \cite{klein2022b}, will be straightforward, as well as the application in finite-element analysis \cite{franke} and optimization of microstructured materials \cite{ortigosa}.
Furthermore, the application to real-world experimental data of composites or polymer materials with varying constituents is targeted.  

%   for PAPER
\vspace*{3ex}

\noindent
\textbf{Conflict of interest.} The authors declare that they have no conflict of interest.
\vspace*{1ex}

\noindent
\textbf{Funding Statement.} This research was supported by the Deutsche Forschungsgemeinschaft (DFG -- German Research Foundation) -- Grant No. 492770117 and the Graduate School of Computational Engineering within the Centre of Computational Engineering at TU Darmstadt.
\vspace*{1ex}

\noindent
\textbf{Data availability.}
The authors provide access to the simulation data required to reproduce the results through the public GitHub repository \url{https://github.com/CPShub/sim-data}.

\renewcommand*{\bibfont}{\footnotesize}
\printbibliography
 
\end{document}